\input harvmac.tex
\input epsf.tex
\input amssym
\input ulem.sty
\input amssym
%\usepackage{amsmath}

%\draftmode

\let\includefigures=\iftrue
\let\useblackboard=\iftrue
\newfam\black

\def\figin{\epsfcheck\figin}\def\figins{\epsfcheck\figins}
\def\epsfcheck{\ifx\epsfbox\UnDeFiNeD
\message{(NO epsf.tex, FIGURES WILL BE IGNORED)}
\gdef\figin##1{\vskip2in}\gdef\figins##1{\hskip.5in}% blank space instead
\else\message{(FIGURES WILL BE INCLUDED)}%
\gdef\figin##1{##1}\gdef\figins##1{##1}\fi}
\def\DefWarn#1{}
\def\figinsert{\goodbreak\midinsert}
\def\ifig#1#2#3{\DefWarn#1\xdef#1{fig.~\the\figno}
\writedef{#1\leftbracket fig.\noexpand~\the\figno} %
\figinsert\figin{\centerline{#3}}\medskip\centerline{\vbox{\baselineskip12pt
\advance\hsize by -1truein\noindent\footnotefont{\bf
Fig.~\the\figno:} #2}}
\bigskip\endinsert\global\advance\figno by1}

% TO INCLUDE FIGURES DO AS BELOW
%\ifig\LABEL{  WRITE CAPTION } {\epsfxsize1.5in\epsfbox{FILENAME.eps}}

%Figure Stuff
\includefigures
\message{If you do not have epsf.tex (to include figures),}
\message{change the option at the top of the tex file.}
\input epsf
\def\figin{\epsfcheck\figin}\def\figins{\epsfcheck\figins}
\def\epsfcheck{\ifx\epsfbox\UnDeFiNeD
\message{(NO epsf.tex, FIGURES WILL BE IGNORED)}
\gdef\figin##1{\vskip2in}\gdef\figins##1{\hskip.5in}% blank space instead
\else\message{(FIGURES WILL BE INCLUDED)}%
\gdef\figin##1{##1}\gdef\figins##1{##1}\fi}
\def\DefWarn#1{}
\def\figinsert{\goodbreak\midinsert}
\def\ifig#1#2#3{\DefWarn#1\xdef#1{fig.~\the\figno}
\writedef{#1\leftbracket fig.\noexpand~\the\figno}%
\figinsert\figin{\centerline{#3}}\medskip\centerline{\vbox{
\baselineskip12pt\advance\hsize by -1truein
\noindent\footnotefont{\bf Fig.~\the\figno:} #2}}
%\bigskip
\endinsert\global\advance\figno by1}
%%%
\else
\def\ifig#1#2#3{\xdef#1{fig.~\the\figno}
\writedef{#1\leftbracket fig.\noexpand~\the\figno}%
%\figinsert\figin{\centerline{#3}}\medskip
%\centerline{\vbox{\baselineskip12pt
%\advance\hsize by -1truein\noindent
%\footnotefont{\bf Fig.~\the\figno:} #2}}
%\bigskip\endinsert

\global\advance\figno by1} \fi

\def\figin{\epsfcheck\figin}\def\figins{\epsfcheck\figins}
\def\epsfcheck{\ifx\epsfbox\UnDeFiNeD
\message{(NO epsf.tex, FIGURES WILL BE IGNORED)}
\gdef\figin##1{\vskip2in}\gdef\figins##1{\hskip.5in}% blank space instead
\else\message{(FIGURES WILL BE INCLUDED)}%
\gdef\figin##1{##1}\gdef\figins##1{##1}\fi}
\def\DefWarn#1{}
\def\figinsert{\goodbreak\midinsert}
\def\ifig#1#2#3{\DefWarn#1\xdef#1{fig.~\the\figno}
\writedef{#1\leftbracket fig.\noexpand~\the\figno} %
\figinsert\figin{\centerline{#3}}\medskip\centerline{\vbox{\baselineskip12pt
\advance\hsize by -1truein\noindent\footnotefont{\bf
Fig.~\the\figno:} #2}}
\bigskip\endinsert\global\advance\figno by1}

\def\newsecB#1{\bigskip\noindent{\bf #1}\writetoca{{#1}}\par\nobreak\medskip\nobreak}

\def \pa {\partial}

\def \la {\langle}
\def \ra {\rangle}
\def \p {\partial}

\def \eps {\epsilon}
\def \b {\bar}

\def \e {\epsilon}

%\KomargodskiEK
\lref\KomargodskiEK{
  Z.~Komargodski and A.~Zhiboedov,
  ``Convexity and Liberation at Large Spin,''
JHEP {\bf 1311}, 140 (2013).
[arXiv:1212.4103 [hep-th]].
%%CITATION = arXiv:1212.4103%%
}

%\FitzpatrickYX
\lref\FitzpatrickYX{
  A.~L.~Fitzpatrick, J.~Kaplan, D.~Poland and D.~Simmons-Duffin,
  ``The Analytic Bootstrap and AdS Superhorizon Locality,''
JHEP {\bf 1312}, 004 (2013).
[arXiv:1212.3616 [hep-th]].
%%CITATION = arXiv:1212.3616%%
}

%\HellermanNRA
\lref\HellermanNRA{
  S.~Hellerman, D.~Orlando, S.~Reffert and M.~Watanabe,
  ``On the CFT Operator Spectrum at Large Global Charge,''
JHEP {\bf 1512}, 071 (2015).
[arXiv:1505.01537 [hep-th]].
%%CITATION = arXiv:1505.01537%%
}

%\MoninJMO
\lref\MoninJMO{
  A.~Monin, D.~Pirtskhalava, R.~Rattazzi and F.~K.~Seibold,
  ``Semiclassics, Goldstone Bosons and CFT data,''
JHEP {\bf 1706}, 011 (2017).
[arXiv:1611.02912 [hep-th]].
%%CITATION = arXiv:1611.02912%%
}

%\ColemanSM
\lref\ColemanSM{
  S.~R.~Coleman, J.~Wess and B.~Zumino,
  ``Structure of phenomenological Lagrangians. 1.,''
Phys.\ Rev.\  {\bf 177}, 2239 (1969).
}

%\CallanSN
\lref\CallanSN{
  C.~G.~Callan, Jr., S.~R.~Coleman, J.~Wess and B.~Zumino,
  ``Structure of phenomenological Lagrangians. 2.,''
Phys.\ Rev.\  {\bf 177}, 2247 (1969).
}

%\MoninBWF
\lref\MoninBWF{
  A.~Monin,
  ``Partition function on spheres: How to use zeta function regularization,''
Phys.\ Rev.\ D {\bf 94}, no. 8, 085013 (2016).
[arXiv:1607.06493 [hep-th]].
%%CITATION = arXiv:1607.06493%%
}

%\LashkariVGJ
\lref\LashkariVGJ{
  N.~Lashkari, A.~Dymarsky and H.~Liu,
  ``Eigenstate Thermalization Hypothesis in Conformal Field Theory,''
[arXiv:1610.00302 [hep-th]].
%%CITATION = arXiv:1610.00302%%
}

%\MaldacenaXJA
\lref\MaldacenaXJA{
  J.~Maldacena and L.~Susskind,
  ``Cool horizons for entangled black holes,''
Fortsch.\ Phys.\  {\bf 61}, 781 (2013).
[arXiv:1306.0533 [hep-th]].
%%CITATION = arXiv:1306.0533%%
}

%\KovtunKW
\lref\KovtunKW{
  P.~Kovtun and A.~Ritz,
  ``Black holes and universality classes of critical points,''
Phys.\ Rev.\ Lett.\  {\bf 100}, 171606 (2008).
[arXiv:0801.2785 [hep-th]].
%%CITATION = arXiv:0801.2785%%
}

%\FreedmanTZ
\lref\FreedmanTZ{
  D.~Z.~Freedman, S.~D.~Mathur, A.~Matusis and L.~Rastelli,
  ``Correlation functions in the CFT(d) / AdS(d+1) correspondence,''
Nucl.\ Phys.\ B {\bf 546}, 96 (1999).
[hep-th/9804058].
%%CITATION = hep-th/9804058%%
}

%\JohnsonASF
\lref\JohnsonASF{
  C.~V.~Johnson,
  ``Critical Black Holes in a Large Charge Limit,''
[arXiv:1705.01154 [hep-th]].
%%CITATION = arXiv:1705.01154%%
}

%\LiRFA
\lref\LiRFA{
  D.~Li, D.~Meltzer and D.~Poland,
  ``Non-Abelian Binding Energies from the Lightcone Bootstrap,''
JHEP {\bf 1602}, 149 (2016).
[arXiv:1510.07044 [hep-th]].
%%CITATION = arXiv:1510.07044%%
}

%\BanerjeeFCX
\lref\BanerjeeFCX{
  D.~Banerjee, S.~Chandrasekharan and D.~Orlando,
  ``Conformal dimensions via large charge expansion,''
[arXiv:1707.00711 [hep-lat]].
%%CITATION = arXiv:1707.00711%%
}

%\RychkovTPC
\lref\RychkovTPC{
  S.~Rychkov and J.~Qiao,
  ``Cut-touching linear functionals in the conformal bootstrap,''
JHEP {\bf 1706}, 076 (2017).
[arXiv:1705.01357 [hep-th]].
%%CITATION = CERN-PH-TH-2017-098%%
}

%\DolanUT
\lref\DolanUT{
  F.~A.~Dolan and H.~Osborn,
  ``Conformal four point functions and the operator product expansion,''
Nucl.\ Phys.\ B {\bf 599}, 459 (2001).
[hep-th/0011040].
%%CITATION = hep-th/0011040%%
}

%\Caron
\lref\Caron{
  S.~Caron-Huot,
  ``Analyticity in Spin in Conformal Theories,''
[arXiv:1703.00278 [hep-th]].
%%CITATION = arXiv:1703.00278%%
}

%\DolanHV
\lref\DolanHV{
  F.~A.~Dolan and H.~Osborn,
  ``Conformal partial waves and the operator product expansion,''
Nucl.\ Phys.\ B {\bf 678}, 491 (2004).
[hep-th/0309180].
%%CITATION = DAMTP-03-91%%
}

%\DolanDV
\lref\DolanDV{
  F.~A.~Dolan and H.~Osborn,
  ``Conformal Partial Waves: Further Mathematical Results,''
[arXiv:1108.6194 [hep-th]].
%%CITATION = arXiv:1108.6194%%
}

%\PappadopuloJK
\lref\PappadopuloJK{
  D.~Pappadopulo, S.~Rychkov, J.~Espin and R.~Rattazzi,
  ``OPE Convergence in Conformal Field Theory,''
Phys.\ Rev.\ D {\bf 86}, 105043 (2012).
[arXiv:1208.6449 [hep-th]].
%%CITATION = LPTENS-12-31%%
}

%\CostaMG
\lref\CostaMG{
  M.~S.~Costa, J.~Penedones, D.~Poland and S.~Rychkov,
  ``Spinning Conformal Correlators,''
JHEP {\bf 1111}, 071 (2011).
[arXiv:1107.3554 [hep-th]].
%%CITATION = LPTENS-11-22%%
}

%\CostaDW
\lref\CostaDW{
  M.~S.~Costa, J.~Penedones, D.~Poland and S.~Rychkov,
  ``Spinning Conformal Blocks,''
JHEP {\bf 1111}, 154 (2011).
[arXiv:1109.6321 [hep-th]].
%%CITATION = LPTENS-11-37%%
}

%\HogervorstSMA
\lref\HogervorstSMA{
  M.~Hogervorst and S.~Rychkov,
  ``Radial Coordinates for Conformal Blocks,''
Phys.\ Rev.\ D {\bf 87}, 106004 (2013).
[arXiv:1303.1111 [hep-th]].
%%CITATION = CERN-PH-TH-2013-043%%
}

%\HartnollAPF
\lref\HartnollAPF{
S.~A.~Hartnoll, A.~Lucas and S.~Sachdev,
``Holographic quantum matter,'' \hfil\break
[arXiv:1612.07324 [hep-th]].
%%CITATION = arXiv:1612.07324%%
}

%\ElShowkHT
\lref\ElShowkHT{
  S.~El-Showk, M.~F.~Paulos, D.~Poland, S.~Rychkov, D.~Simmons-Duffin and A.~Vichi,
  ``Solving the 3D Ising Model with the Conformal Bootstrap,''
Phys.\ Rev.\ D {\bf 86}, 025022 (2012).
[arXiv:1203.6064 [hep-th]].
%%CITATION = LPTENS-12-07%%
}

%\Simmons-DuffinWLQ
\lref\SimmonsDuffinWLQ{
  D.~Simmons-Duffin,
  ``The Lightcone Bootstrap and the Spectrum of the 3d Ising CFT,''
JHEP {\bf 1703}, 086 (2017).
[arXiv:1612.08471 [hep-th]].
%%CITATION = arXiv:1612.08471%%
}

%\CardyIE
\lref\CardyIE{
  J.~L.~Cardy,
  ``Operator Content of Two-Dimensional Conformally Invariant Theories,''
Nucl.\ Phys.\ B {\bf 270}, 186 (1986)..
}

%\HartmanOAA
\lref\HartmanOAA{
  T.~Hartman, C.~A.~Keller and B.~Stoica,
  ``Universal Spectrum of 2d Conformal Field Theory in the Large c Limit,''
JHEP {\bf 1409}, 118 (2014).
[arXiv:1405.5137 [hep-th]].
%%CITATION = CALT-68-2889%%
}

%\PelissettoEK
\lref\PelissettoEK{
  A.~Pelissetto and E.~Vicari,
  ``Critical phenomena and renormalization group theory,''
Phys.\ Rept.\  {\bf 368}, 549 (2002).
[cond-mat/0012164].
%%CITATION = cond-mat/0012164%%
}

%\HeemskerkPN
\lref\HeemskerkPN{
  I.~Heemskerk, J.~Penedones, J.~Polchinski and J.~Sully,
  ``Holography from Conformal Field Theory,''
JHEP {\bf 0910}, 079 (2009).
[arXiv:0907.0151 [hep-th]].
%%CITATION = arXiv:0907.0151%%
}

%\HellermanVEG
\lref\HellermanVEG{
  S.~Hellerman, S.~Maeda and M.~Watanabe,
  ``Operator Dimensions from Moduli,''
[arXiv:1706.05743 [hep-th]].
%%CITATION = IPMU-17-0015%%
}

%\Alvarez-GaumeVFF
\lref\AlvarezGaumeVFF{
  L.~Alvarez-Gaume, O.~Loukas, D.~Orlando and S.~Reffert,
  ``Compensating strong coupling with large charge,''
JHEP {\bf 1704}, 059 (2017).
[arXiv:1610.04495 [hep-th]].
%%CITATION = CERN-TH-2016-221%%
}

%\HellermanEFX
\lref\HellermanEFX{
  S.~Hellerman, N.~Kobayashi, S.~Maeda and M.~Watanabe,
  ``A Note on Inhomogeneous Ground States at Large Global Charge,''
[arXiv:1705.05825 [hep-th]].
%%CITATION = IPMU17-0081%%
}

%\BanerjeeFCX
\lref\BanerjeeFCX{
  D.~Banerjee, S.~Chandrasekharan and D.~Orlando,
  ``Conformal dimensions via large charge expansion,''
[arXiv:1707.00711 [hep-lat]].
%%CITATION = arXiv:1707.00711%%
}

%\HartmanOAA
\lref\HartmanOAA{
  T.~Hartman, C.~A.~Keller and B.~Stoica,
  ``Universal Spectrum of 2d Conformal Field Theory in the Large c Limit,''
JHEP {\bf 1409}, 118 (2014).
[arXiv:1405.5137 [hep-th]].
%%CITATION = CALT-68-2889%%
}

%\FitzpatrickZHA
\lref\FitzpatrickZHA{
  A.~L.~Fitzpatrick, J.~Kaplan and M.~T.~Walters,
  ``Virasoro Conformal Blocks and Thermality from Classical Background Fields,''
JHEP {\bf 1511}, 200 (2015).
[arXiv:1501.05315 [hep-th]].
%%CITATION = arXiv:1501.05315%%
}

%\LoukasLOF
\lref\LoukasLOF{
  O.~Loukas, D.~Orlando and S.~Reffert,
   ``Matrix models at large charge,'' \hfil\break
[arXiv:1707.00710 [hep-th]].
%%CITATION = arXiv:1707.00710%%
}

%\Hasenbusch
\lref\Hasenbusch{
  M.~Hasenbusch and E.~Vicari,
  ``Anisotropic perturbations in three-dimensional $O(N)$-symmetric vector models,''
  Phys. Rev. B {\bf 84}, 125136
[arXiv:1108.0491 [cond-mat.stat-mech]].
%%CITATION = arXiv:1707.00710%%
}

%\QiaoXIF
\lref\QiaoXIF{
  J.~Qiao and S.~Rychkov,
  ``A tauberian theorem for the conformal bootstrap,''
[arXiv:1709.00008 [hep-th]].
%%CITATION = CERN-TH-2017-176%%
}

%\ArkaniHamedDZ
\lref\ArkaniHamedDZ{
  N.~Arkani-Hamed, L.~Motl, A.~Nicolis and C.~Vafa,
  ``The String landscape, black holes and gravity as the weakest force,''
JHEP {\bf 0706}, 060 (2007).
[hep-th/0601001].
%%CITATION = hep-th/0601001%%
}

%\HellermanSUR
\lref\HellermanSUR{
  S.~Hellerman and S.~Maeda,
  ``On the Large $R$-charge Expansion in ${\cal N} = 2$ Superconformal Field Theories,''
[arXiv:1710.07336 [hep-th]].
%%CITATION = arXiv:1710.07336%%
}

%\AldayEYA
\lref\AldayEYA{
  L.~F.~Alday, A.~Bissi and T.~Lukowski,
  ``Large spin systematics in CFT,''
JHEP {\bf 1511}, 101 (2015).
[arXiv:1502.07707 [hep-th]].
%%CITATION = arXiv:1502.07707%%
}

%\FitzpatrickZHA
\lref\FitzpatrickZHA{
  A.~L.~Fitzpatrick, J.~Kaplan and M.~T.~Walters,
  ``Virasoro Conformal Blocks and Thermality from Classical Background Fields,''
JHEP {\bf 1511}, 200 (2015).
[arXiv:1501.05315 [hep-th]].
%%CITATION = arXiv:1501.05315%%
}

%\FitzpatrickZHA
\lref\Newton{
https://en.wikipedia.org/wiki/Newton's\_identities
}

\lref\Deutsch{
J.~M.~Deutsch, 
``Quantum statistical mechanics in a closed system," 
Physical Review A, {\bf 43(4)}, 2046 (1991). 
}

\lref\Srednicki{
M.~Srednicki,
``Chaos and quantum thermalization,"
Physical Review E, {\bf 50(2)}, 888 (1994).
[cond-mat/9403051].
}

\lref\Rigol{
M.~Rigol, V.~Dunjko and M.~Olshanii, 
``Thermalization and its mechanism for generic isolated quantum systems," 
Nature, {\bf 452(7189)}, 854-858 (2008).
[arXiv:0708.1324 [cond-mat.stat-mech]].
}

%\DAlessioRWT
\lref\DAlessioRWT{
  L.~D'Alessio, Y.~Kafri, A.~Polkovnikov and M.~Rigol,
  ``From quantum chaos and eigenstate thermalization to statistical mechanics and thermodynamics,''
Adv.\ Phys.\  {\bf 65}, no. 3, 239 (2016).
[arXiv:1509.06411 [cond-mat.stat-mech]].
%%CITATION = arXiv:1509.06411%%
}

\Title{
\vbox{\baselineskip8pt
% \hbox{SPIN-07/41} \hbox{
%ITP-UU-07/55}
}}
{\vbox{
\centerline{Conformal Bootstrap At Large Charge}
}}

\bigskip
\centerline{Daniel Jafferis, Baur Mukhametzhanov and Alexander Zhiboedov}
\bigskip
\centerline{\it Department of Physics, Harvard University, Cambridge, MA 20138, USA
}

\vskip .5in

\noindent
We consider unitary CFTs with continuous global symmetries in $d>2$. We consider a state created by the lightest operator of large charge $Q \gg 1$ and analyze the correlator of two light charged operators in this  state. We assume that the correlator admits a well-defined large $Q$ expansion and, relatedly, that the macroscopic (thermodynamic) limit of the correlator exists. We find that the crossing equations admit a consistent truncation, where only a finite number $N$ of Regge trajectories contribute to the correlator at leading nontrivial order. We classify all such truncated solutions to the crossing. For one Regge trajectory $N=1$, the solution is unique and given by the effective field theory of a Goldstone mode. For two or more Regge trajectories $N \geq 2$, the solutions are encoded in roots of a certain degree $N$ polynomial. Some of the solutions admit a simple weakly coupled EFT description, whereas others do not. In the weakly coupled case, each Regge trajectory corresponds to a field in the effective Lagrangian.

\Date{}

%\listtoc\writetoc
%\vskip .5in \noindent

\listtoc\writetoc
\vskip 0.7in \noindent

\break

\newsec{Introduction}
In this paper we consider CFTs in $d>2$ with continuous global symmetries. The spectrum of these CFTs contains operators charged under these symmetries. For simplicity, we focus on the case of $U(1)$. We denote the lightest operator of charge $Q$ as ${\cal O}_{Q}$, its dimension being $\Delta_Q$. We are interested in the limit when $Q$ becomes large.

One of the simplest nontrivial examples of this type is given by the $O(2)$ Wilson-Fischer CFT in $d=3$. This theory has $U(1) \times {\Bbb Z}_2$ global symmetry and is common in Nature, see, e.g., \PelissettoEK. It is commonly defined as the IR fixed point of the flow generated by the $(\phi^{\dagger} \phi)^2$ deformation of the free complex scalar theory in the UV. In a recent paper \HellermanNRA\ it was argued that the large $Q$ subsector of this theory is described by a conformally invariant effective field theory (EFT) Lagrangian of a Goldstone boson. In particular, the authors of \HellermanNRA\ predicted the spectrum of operators $\Delta$ with dimensions slightly above $\Delta_Q$, namely the operators with $\Delta - \Delta_Q \sim O(1)$ in the large $Q$ limit. This approach was further developed in \MoninJMO, where the correlation functions of light charged operators in the background of the heavy state were computed. Generalizations to systems with more symmetries were found in \refs{\MoninJMO\AlvarezGaumeVFF\HellermanEFX-\LoukasLOF}. Some of the EFT predictions have been tested using Monte-Carlo simulations in \Hasenbusch, \BanerjeeFCX.

These results are the starting point for our analysis. We would like to understand how universal they are and what assumptions would go into their derivation in generic CFTs. Therefore, we study a crossing equation for heavy-heavy-light-light operators in an abstract CFT with a global symmetry. We take the heavy state to be ${\cal O}_{Q}$, the lightest operator with a given large charge. Notice that the large $Q$ limit is different from the more familiar large $N_c$ \HeemskerkPN,  or large spin $J$ limits  \FitzpatrickYX, \KomargodskiEK. In the latter cases one considers a fixed correlator and changes either parameters of the theory or cross ratios within the correlator. In the case of large $Q$ we analyze the limit of a family of correlators within one theory. Indeed, for every $Q$ the external operator is different. This leads to several peculiarities in the analysis of the crossing equation that we will discuss below. Nevertheless, we assume that correlation functions that involve ${\cal O}_{Q}$ admit a smooth large $Q$ limit, namely that we can build an expansion in inverse powers of $Q$ and think of $Q$ as a smooth parameter.

The essential simplification of the large $Q$ limit is that in a certain domain in the space of cross ratios, the dominant contributions to the four-point function in the heavy-light fusion channels come from a set of operators whose dimensions above that of the lightest large charge operator are of order $1$ in the $Q$ scaling. These are the operators that are characterized by the effective field theory.

We begin by performing a detailed analysis of crossing for the four-point function computed from  effective field theory in \MoninJMO. This sets the stage for a more abstract analysis. One immediate feature of the large $Q$ limit is that in the $z$ conformal frame, the contribution of the descendants to the conformal blocks are suppressed by a power of $Q$. This greatly simplifies the problem. For example, at leading order crossing is satisfied by a single operator! Further analysis, however, reveals two discomforting features:

$a)$ $s$- and $u$-channel (heavy-light) OPEs do not have an overlapping region of convergence within EFT;\foot{In the full CFT the $s$- and $u$- channels, of course, converge as always. However, in the overlapping region, the dominant operators in one of the channels are not the ones that are described by the EFT.}

$b)$ when EFT is applicable, the $t$-channel (light-light) OPE is dominated by unknown neutral heavy operators.

\noindent This looks like an impasse for any bootstrap analysis. There is, however, the third feature of the EFT result which allows us to make a further progress:

$c)$ at leading nontrivial order only one Regge trajectory contributes to the $s$- and $u$- channel OPEs.

The presence of a single Regge trajectory in the conformal block decomposition of the EFT result is a direct consequence of having only one field, the Goldstone mode, in the EFT Lagrangian. A priori, it is not obvious that crossing equations admit solutions with only a finite number of Regge trajectories. Indeed, without taking the large $Q$ limit this would be impossible, and it is a nontrivial property of the conformal blocks in the large $Q$ limit.

It is, therefore, natural to consider a truncated ansatz for the correlation function, where only a finite number of Regge trajectories contribute to the correlator at the first nontrivial order in the $s$- and $u$-channel OPEs. This seems to be a weak CFT version of what we mean by having an effective field theory description of the large $Q$ correlators. Moreover, we impose that this ansatz satisfies the following properties:

$a')$ Smooth matching of $s$- and $u$- channels at their common boundary of convergence.

$b')$ Existence of the macroscopic (thermodynamic) limit of the correlator.

\noindent In $a')$, instead of the $s=u$ crossing we impose analyticity of the correlation function at the boundary of convergence of the large $Q$ limit of each channel, namely that the two expansions should match smoothly. This argument is similar in spirit to the one used in \PappadopuloJK, where the asymptotic density of operators and three-point functions were found. In our case it becomes much more powerful, due to the crucial assumption, motivated by EFT considerations, that only $N$ Regge trajectories are present in the OPE at this order. This is a weak CFT version of a notion of having a finite number of ``fields'' in the ``EFT.''

In $b')$, we note that in the absence of a controlled $t$-channel OPE, the short distance behavior of the correlator is controlled by the existence of what we will call a {\it macroscopic limit}. This limit was recently discussed in  \LashkariVGJ\ in the context of the eigenstate thermalization hypothesis (ETH) \refs{\Deutsch\Srednicki-\Rigol} (for a review, see, e.g., \DAlessioRWT). Consider a CFT state on a cylinder ${\Bbb R} \times S^{d-1}$, and take the radius of the sphere $R \to \infty$ while keeping the correlators of light operators finite by appropriately scaling the parameters of the state, in this case its charge and energy.  Equivalently, this is a combined limit in which the scaling dimension of the external operator is taken to infinity as we tune cross ratios appropriately.

The existence of such limits, which result in flat space correlators in a nontrivial background, seems to be a generic feature of any CFT. Typically, the energy and charge density of the state will remain fixed when the limit is taken such that the correlators remain finite.\foot{One also needs to rescale the light operators appropriately. The nontrivial condition is that this can be done to keep all (2 heavy $+ n$ light)-point functions finite.} An important exception that we will discuss further in section 5 appears when there is a moduli space of vacua. We assume that such a limit exists. This type of limit seems to be a generic feature of any CFT, and indeed it exists for the case analyzed in \HellermanNRA, \MoninJMO.

Furthermore, for generic heavy operators, that are not the lightest carrying some large charge, the physics of the macroscopic limit is expected to be thermal, and described by hydrodynamics at finite temperature. However, in the situation at hand, for the lightest operators with large charge, we expect a finite charge density configuration at zero temperature, associated with the quantum EFT, in the macroscopic flat space limit.

Under the assumptions\ $a')$, $b')$, $c)$ we classify the leading order solutions to the crossing equation. The solutions for scaling dimensions as functions of spin are given by the roots of a certain polynomial that we describe in detail below. For $N=1$ we show that the Goldstone EFT is the unique solution. For $N\geq 2$ there are many possibilities. Some of them correspond to adding extra particles. Other solutions do not come from any weakly coupled EFT Lagrangian. At present, we do not know which of the solutions are realized in CFTs and could be consistently promoted to a solution of the crossing equations higher orders in ${1 \over Q}$. We leave these questions for the future.

In section 2 we describe general features of the large $Q$ limit. In section 3 we describe the basic kinematics and the properties of conformal blocks in the large $Q$ limit. In section 4 we review the results of EFT for the spectrum and the four-point function. In section 5 we describe the macroscopic limits in a generic CFT. In section 6 we perform the bootstrap analysis of the four-point function. In section 7 we present some extensions by considering operators with spin, going to next order and doing an analog of the light-cone bootstrap in the macroscopic limit.
We end with conclusions and future directions.

\newsec{Large $Q$ Limit}

\subseclab\largeQ

We will be interested in the large charge $Q$ limit of correlators of the type
\eqn\LargeCh{
G^{Q}(x_i)  \equiv \la {\cal O}_{Q}(0) {\cal O}_{q_1} (x_1) \dots {\cal O}_{q_n}(x_n) {\cal O}_{q_{n+1}} (1) {\cal O}_{-Q}(\infty) \ra \ ,
}
where we used conformal symmetry to fix the positions of three operators.

Note that the large $Q$ limit is taken not within a fixed correlator, but rather it is a limit of a family of different correlators which involve different operators ${\cal O}_{Q}$. Charge $Q$ being a discrete quantum number, one might wonder to what extent this limit is well-defined. A somewhat similar situation arises in the discussion of the large spin $J$ limit \FitzpatrickYX, \KomargodskiEK. There, however, one can argue \Caron\ that the CFT data is analytic in spin. A core assumption of the present work is that a similar analyticity exists in charge as well. Namely, we will treat operators with large $Q$ and their corresponding three-point functions as smooth functions of $Q$ that admit a large $Q$ expansion.

Imagine now a family of operators labeled by $Q$. The discreteness of the spectrum implies that
\eqn\condition{
\lim_{Q \to \infty} \Delta_{min}(Q) \to \infty \ .
}
Indeed, otherwise we would have an infinite number of operators with bounded dimensions $\Delta \leq \Delta^*$.

As we make $Q$ large, there are several possibilities. If the spectrum close to the lightest state is sparse, the correlator \LargeCh\ is dominated by the ``vacuum'' in each OPE channel together with excitations with energy of $O(1)$, namely
\eqn\LargeCh{
G^{Q}(z_i, \bar z_i )  \sim e^{- \sum_{i=1}^{n} \Delta_{min} (Q + \sum_{k=1}^{i} q_k)|\tau_{i+1} - \tau_i|} \ ,
}
where the proportionality coefficient is given by the corresponding three-point couplings. At this point we assume a type of large $Q$ clustering, namely that there is no extra large Euclidean time scale at which the picture \LargeCh\ breaks down. In principle, one could imagine a slightly heavier operator with an enhanced three-point function, which would lead to an extra factor of $Q^{\alpha} e^{- \beta \delta \tau}$ in comparison with \LargeCh. Then for $\delta \tau \gg \log Q$ it would be suppressed, whereas for $\delta \tau \ll \log Q$ it would be dominant. We assume that this does not happen and the same lightest state dominates the correlator at large Euclidean times.

There are other possibilities that we do not consider. For example, one can imagine that there is a parametrically large degeneracy of operators close to the lightest state. This would be the case, for example, if the dual state were an extremal Reissner-Nordstr\"{o}m black hole. It would be interesting to study this possibility or rule it out.

For simplicity, let us set $n=1$ and $q_1 = -q_2 = - q$, so we are considering a four-point function. We have
\eqn\LargeChB{
G^{Q}(z, \bar z )  \sim \lambda_{\Delta_{min}(Q-q)}^2 (z \bar z)^{{\Delta_{min}(Q-q) \over 2}} \ .
}
The minimal energy state could be degenerate or carry spin. The large $Q$ expansion is dominated by the minimal energy state and $O(1)$ energy excitations around it. Note that operators that are parametrically heavier than $\Delta_{min}$ are non-perturbatively suppressed by the factor $(z \bar z)^{(\Delta_* - \Delta_{min})/2}$. This picture is very similar to the usual saddle point approximation with $Q$ playing the role of ${1 \over \hbar} \to \infty$. Operators with scaling dimensions parametrically different in $Q$ correspond to different saddles, whereas the fluctuations around a given saddle are described by the operators which have parametrically the same scaling dimensions.

A state on the cylinder created by the lightest operator of charge $Q$ is characterized by the energy density $\eps$ and charge density $q$
\eqn\energyandchargedensityA{
\eps = {\Delta_Q \over R^{d}}, ~~~ q = {Q \over R^{d-1}} \ .
}
where we used the fact that $E_{cyl} = {\Delta_Q \over R}$. As we take $Q$ to be large we can simultaneously take $R \to \infty$ so that $\eps$ is kept fixed. Generically, we expect that finite charge density $q$ states carry some fixed non-zero energy density as well. This implies that
\eqn\largeQdim{
\Delta_{min}(Q) \equiv \Delta_Q \sim Q^{d/(d-1)} \ .
}
We expect \largeQdim\ to hold in generic interacting CFTs.\foot{The situation is different in CFTs with a nontrivial moduli space of vacua. We discuss this in more detail in section 5.} Alternatively, \largeQdim\ is a consequence of a local relationship between charge and energy densities \HellermanNRA. In the present work we mostly focus on the case \largeQdim, except for some parts of section 5 and section 4.7.

\newsec{Four-point Function Kinematics}

 In this section we review basic kinematics of the four-point correlator and set our conventions. We consider a four-point function of scalar operators
\eqn\fourpoint{\eqalign{
G (z,\bar z) &\equiv\la {\cal O}_{Q}(0) {\cal O}_{-q} (z, \bar z) {\cal O}_{q} (1) {\cal O}_{-Q}(\infty) \ra \ , \cr
u &= {x_{12}^2 x_{34}^2 \over x_{13}^2 x_{24}^2}= z \bar z, ~~~ v = {x_{14}^2 x_{23}^2 \over x_{13}^2 x_{24}^2} = (1-z)(1-\bar z) \ ,
}}
where the charge $Q$ is very large and ${\cal O}_Q$ has the smallest dimension $\Delta_Q$ among operators with charge $Q$. As discussed in the previous section, as we take $Q \to \infty$, we have $\Delta_Q \to \infty$ as well. Therefore, \fourpoint\ describes a heavy-light-light-heavy correlation function.\foot{For a related discussion in $d=2$, see \FitzpatrickZHA.} It is instructive to analyze what happens to the conformal blocks in this limit.

The correlator $G(z, \bar z)$ admits an expansion in terms of conformal blocks in three different channels
\eqn\blockss{\eqalign{
{\rm s-channel:} ~~~G(z,\bar z) = (z \bar z)^{- {1 \over 2} (\Delta_Q + \Delta_{q}) } \sum_{{\cal O}_{\Delta, J}} | \lambda_{Q,-q, {\cal O}_{\Delta, J}} |^2 g_{\Delta, J}^{\Delta_{Q,q}, - \Delta_{Q,q}}(z, \bar z) \ , \cr
~~~ \Delta_{Q,q} = \Delta_{Q} - \Delta_{q},~~~ |z|<1 \ .
}}
\eqn\blockst{\eqalign{
{\rm t-channel:} ~~~G(z,\bar z) &= \left([1-z][1- \bar z]\right)^{- \Delta_q } \cr
&\sum_{ {\cal O}_{\Delta, J} } \lambda_{-q,q, {\cal O}_{\Delta, J}} \lambda_{Q,-Q , {\cal O}_{\Delta, J}} g_{\Delta, J}^{0,0}(1-z,1-\bar z) ,\quad   |1-z|<1 \ .
}}
\eqn\blocksu{\eqalign{
{\rm u-channel:} ~~~G(z,\bar z) = (z \bar z)^{{1 \over 2} (\Delta_Q - \Delta_{q}) } \sum_{ {\cal O}_{\Delta, J} } | \lambda_{Q,q, {\cal O}_{\Delta, J}} |^2 g_{\Delta, J}^{\Delta_{Q,q}, - \Delta_{Q,q}}\left({1 \over z}, {1 \over \bar z} \right) \ , \cr
~~~ |z|>1 \ ,\cr
}}
where the sum is over an infinite set of primary operators that appear in the corresponding OPE channel. The expressions for conformal blocks can be found, for example, in \DolanDV.

It is also convenient to define $g_q(z,\b{z})$ as
\eqn\gqdef{
g_q(z,\b{z}) \equiv (z\b{z})^{{1\over 2} \Delta_q}  G(z,\b{z}) \ .
}

The correlation function is invariant if we send $q$ to $-q$ and exchange the locations of the two light operators, ${\cal O}_{-q} \leftrightarrow {\cal O}_q$. This is encoded in the crossing equation $s=u$
\eqn\crossingequation{
g_q(z, \bar z) =  g_{-q} \left({1 \over z}, {1 \over \bar z} \right) \ .
}

Of course, finding the most generic $G(z, \bar z) $ consistent with the OPE and crossing is an insurmountable task. The key simplification here is that there is a small parameter in the problem, namely $Q^{-1} \ll 1$. This allows us to find some universal features in the limit.

We will also need correlation functions on a cylinder ${\Bbb R} \times S^{d-1}$, which is conformally mapped to the plane ${\Bbb R}^d$ by $(\tau, {\bf n}) \to (r = R e^{\tau}, {\bf n})$. In the conformal frame \fourpoint, when all operators lie in the same plane, we have a relation between cylinder coordinates and $z,\b z$ plane coordinates
\eqn\map{\eqalign{
&z = e^{\tau + i \theta}, \qquad \b z = e^{\tau  - i \theta} \ , \cr
&z\b{z} = e^{2 \tau}, \qquad {z + \b{z} \over 2\sqrt{z \b{z}}} = \cos \theta \equiv x \ ,
}}
where $\theta$ is the angle between two light operators in \fourpoint. Primary operators transform as
\eqn\mapop{
{\cal O}_{cyl}(\tau,{\bf n}) = \left( r \over R \right)^{\Delta_{\cal O}} {\cal O}(r, {\bf n}) \ .
}

As a prerequisite for studying bootstrap equations, we review the structure of conformal blocks in the large charge limit.

\subsec{Conformal Blocks in The Large $Q$ Limit}

As is evident from \blockss, \blockst, \blocksu, conformal blocks depend on $Q$ only in the $s$- and $u$-channels. To understand the structure of the blocks it is instructive to write them as a sum over descendants
\eqn\confmblock{
g^{\Delta_{Q,q};-\Delta_{Q,q}}_{\Delta_{Q-q},J}(z,\b{z}) = \sum_{n=0}^{\infty} \sum_j a_{j,n}(z\b{z})^{\Delta_{Q-q} + n  \over 2} C_j^{({d \over 2}-1)} \left(z + \b{z} \over 2\sqrt{z\b{z}} \right) \ ,
}
and fix all the coefficients $a_{j,n}$ by solving the Casimir equation; here $C_J^{({d \over 2} -1)}(x)$ are the usual Gegenbauer polynomials which become Legendre polynomials in $d=3$. Say, for $n=0$ we have $j=J$; for $n=1$ we have $j = J+1$ and $j = J-1$, etc. An explicit solution for $a_{j,n}$'s in general case was found in \DolanHV .
Let us write explicitly the first nontrivial correction due to the level one descendants
\eqn\blockQ{\eqalign{
g^{\Delta_{Q,q};-\Delta_{Q,q}}_{\Delta_{Q-q},J}(z,\b{z}) &=  (z\b{z})^{\Delta_{Q-q}  \over 2} \left( C_J^{({d \over 2}-1)}+ \sqrt{z \bar z} \left( a_{J+1,1}  C_{J+1}^{({d \over 2}-1)} + a_{J-1,1}  C_{J-1}^{({d \over 2}-1)}\right) + O(z \bar z) \right) ,\cr
a_{J+1,1} &= {1 \over 2 (d- 2 + 2 J)} {(J+1)(\Delta_{Q-q} - \Delta_{Q,q} + J)^2 \over \Delta_{Q-q} + J} \ , \cr
a_{J-1,1} &= {1 \over 2 (d- 2 + 2 J)}  {(J+d-3)(\Delta_{Q-q} - \Delta_{Q,q}-J-d+2)^2 \over \Delta_{Q-q} - J - d +2} \ ,
}}
where for the sake of brevity we omitted the arguments of Gegenbauer polynomials which are the same as in \confmblock. When $J=0$, the term $C_{J-1}^{({d \over 2}-1)}$ is absent. For our purposes we will also need the contribution of the level-two descendants in the case $J=0$, which take the form
\eqn\leveltwoscalar{\eqalign{
a_{2,2} &= {1 \over 4 d (d- 2)} {(\Delta_{Q-q} - \Delta_{Q,q} )^2 (\Delta_{Q-q} - \Delta_{Q,q} +2 )^2\over \Delta_{Q-q} (\Delta_{Q-q}+1)} \ , \cr
a_{0,2} &={1 \over 4 d}  {(\Delta_{Q-q} - \Delta_{Q,q} )^2 (\Delta_{Q-q} - \Delta_{Q,q} -d+2)^2 \over \Delta_{Q-q} (2 \Delta_{Q-q}  - d +2)} \ .
}}
We would like to consider the limit of $Q \gg 1$ and fixed $d,J$. In the  conformal bootstrap analysis of the large charge EFT, we will be interested in operators $\Delta_{Q-q}$ and $\Delta_Q$ belonging to the same family \largeQdim\ $\Delta_Q \sim Q^{{d \over d-1}}$. It is then clear from \blockQ, \leveltwoscalar\ that the contribution of descendants is governed by the parameter
\eqn\asymptotic{
{\left(\Delta_{Q-q} - \Delta_Q \right)^2 \over \Delta_{Q-q}} \sim {1 \over \Delta_Q} \left({\p \Delta_Q \over \p Q} \right)^2 \sim Q^{- {d - 2 \over d-1}} \to 0 \ .
}
Therefore, for $d>2$ the expansion \confmblock\ is a controlled approximation of the conformal block in the large charge $Q$ limit. To leading order at large $Q$ only primary operators contribute. This simplifies our analysis in later sections.

One can use recursion relations for Gegenbauer polynomials to simplify \blockQ\ in the large $Q$ limit.
The result for the first subleading correction takes the form
\eqn\blockQsub{
g^{\Delta_{Q,q};-\Delta_{Q,q}}_{\Delta_{Q-q},J}(z,\b{z})=  (z\b{z})^{\Delta_{Q-q}  \over 2} C_J^{({d \over 2}-1)} \left(z + \b{z} \over 2\sqrt{z\b{z}} \right)\left( 1 + {q^2 \over 4} {1 \over \Delta_Q} \left( \p \Delta_Q \over \p Q \right)^2 (z+\b{z}) + ... \right) .
}
Curiously, the first correction in the parentheses does not depend on $J$. We do not have an explanation for this fact beyond direct computation.

%%%%%%%%%%%%%%%%%%%%%%%%%%%%%%%%%%%%%%%%%%%%%%%%%%%%%%%%%%%%%%%%%%%%%%%%%%%%%%

\newsec{Effective Field Theory}

The goal of this section is to provide the reader with the results for the operator spectrum \HellermanNRA\ and correlation functions \MoninJMO\ at large charge and review the tools of effective field theory necessary to obtain them.

We consider a CFT with some global symmetry group $G$ and assume that the CFT spectrum contains operators charged under this symmetry (which implies that there exist operators of arbitrarily large charge $Q$, by repeated OPE contraction). For simplicity, we focus on the case $G = U(1)$.

The essential idea of {\HellermanNRA} is the following. Let us consider an operator of charge $Q$, ${\cal O}_Q$.
By the operator/state correspondence this operator describes a state with  charge density $\rho \sim {Q \over R^{d-1}} $ on the cylinder ${\Bbb R} \times S^{d-1}$. Here $R$ stands for the radius of the sphere.  In the limit $Q \gg 1$ there is a large separation of UV and IR scales $\rho \gg {1\over R^{d-1}}$. One can view the state with charge $Q$ as spontaneously breaking the $U(1)$ symmetry (as well as some of the space-time symmetries \MoninJMO). This leads to the existence of a massless Goldstone boson. At distances much bigger than the distance set by the charge density, this Goldstone  mode is described by an EFT corresponding to a particular symmetry breaking pattern. The expansion parameter in the EFT is the ratio of UV and IR scales ${\rho^{-1/(d-1)} \over R} \sim Q^{-1/(d-1)} \ll 1$.

The state with homogeneous charge density $\rho$ on the cylinder ${\Bbb R} \times S^{d-1}$ breaks the global symmetry group $SO(d+1,1)\times U(1)$ down to rotations of the sphere $SO(d)$ and a linear combination of $U(1)$ and time translations\foot{We put a hat on the $U(1)$ generator $\widehat{Q}$ to distinguish it from the c-number $Q$.} ${\cal H}' = {\cal H} + \mu \widehat{Q}$

\eqn\symmbreak{
SO(d+1,1) \times U(1) \to SO(d) \times {\cal H}' \ .
}
The corresponding effective Lagrangian can be obtained using the CCWZ construction \ColemanSM, \CallanSN. It can be written in terms of a field $\chi(x)$, whose fluctuations around an appropriate saddle describe the Goldstone boson $\pi(x)$.

In particular, in three dimensions $d=3$ we have \refs{\HellermanNRA,\MoninJMO} (in  the Euclidean signature)\foot{In our convention the curvature of $S^n$ is ${\cal R} = {n(n-1) \over R^2}$.}
\eqn\EFTaction{\eqalign{
S_E = & - \int d^3 x \sqrt{g} \Bigg( {1 \over 12\pi \alpha^2} |\p \chi|^3 - {\beta \over 8\pi \alpha} |\p \chi| \left( {\cal R} + 2 {(\p_{\mu} |\p\chi|) (\p^{\mu} |\p\chi|) \over |\p\chi|^2 } \right)  + \dots
... \Bigg) + \cr
& ~~~~ + i \rho \int d^3 x \sqrt{g} \dot{\chi} \ ,
}  }
where $|\p\chi| \equiv (- g^{\mu\nu}\p_\mu \chi \p_\nu \chi)^{1/2}$ and $\alpha, \beta, \gamma$ are undetermined coupling constants of the EFT.\foot{Our definitions of $\alpha, \beta, \gamma$ are related to $c_1,c_2,c_3$ in {\MoninJMO} by $\alpha =  {1\over \sqrt{2\pi c_1} }, \beta = - {8\pi c_2 \over \sqrt{2\pi c_1}}, \gamma = c_3$. This normalization will be more convenient for scaling dimensions and correlation functions.} By ellipsis we denote higher order curvature couplings, which are suppressed by $1 \over Q$ when we expand around the relevant saddle point. This action is Weyl invariant assuming that the metric has Weyl weight two and $\chi$ has Weyl weight zero. The field $\chi$ transforms by shifts under $U(1)$, with the corresponding charge density being $j^0(\tau, {\bf n}) = {\p L \over \p \dot{\chi}}$.\foot{The Lagrangian $L$ here does not include the chemical potential.} The last term in {\EFTaction} is the chemical potential which sets the charge density $j^0(\tau, {\bf n})$ to a constant value $\rho = {Q \over 4 \pi R^2}$. Note that this action is meaningful only when expanded around the saddle described below that gives the large charge state. Therefore it can be regarded as a tool for constructing the Goldstone action by giving a simpler realization of the broken symmetries. In particular, the $\chi$ field is not meaningful near $\chi = 0$, and the above action is not meant to approximate the exact CFT in that regime.

To leading order one can use the first term in {\EFTaction} to obtain the saddle-point. Assuming that the lowest energy state is homogeneous on $S^2$, the saddle-point is simply given by
\eqn\saddle{\eqalign{
\writedefs
\b{\chi} &= -i\mu \tau + \chi_0 \ , \cr
\mu R &= \alpha \sqrt{Q} + {\beta \over 2\sqrt{Q}} + O(Q^{-3/2}) \ ,
}}
where $\mu, \chi_0$ are constants and $\mu$ is fixed by the eom at $\tau = \pm \infty$.\foot{Equivalently, we could have fixed $\mu$ by imposing $\la Q | j^0 | Q \ra = {Q \over 4 \pi R^2}$.} Since $\chi$ transforms by shifts under $U(1)$, this solution indeed preserves ${\cal H}' = {\cal H}+ \mu \widehat{Q}$ in accordance with {\symmbreak}.

Expanding the action {\EFTaction} around the saddle {\saddle}
\eqn\GBfluct{
\chi(x) = -i \mu \tau + {\alpha \over 2} {1\over \sqrt{\mu}}  \pi(x)
}
we find
\eqn\saddleexp{\eqalign{
S_E = &{\Delta_Q \over R} (\tau_{out} - \tau_{in}) + S_{\pi} \ , \cr
S_{\pi} = &{1\over 16 \pi} \int d^3 x \sqrt{g} \left( \dot{\pi}^2 + {1\over 2} (\p_i \pi)^2 \right)+ \cr
&+{i \over 96 \pi \sqrt{\alpha}} {1\over Q^{3/4}} \int d^3x \sqrt{g} \left( \dot \pi ^3 + {3\over 2} \dot \pi (\p_i \pi)^2 \right) +
O(Q^{-1}) \ , \cr
\Delta_Q = &{2\over 3} \alpha  Q^{3/2} + \beta \sqrt{ Q } + C + O(Q^{-1/2}) \ ,
}}
where $E_Q = {\Delta_Q \over R}$ is the energy of the ``vacuum'' state in the large charge $Q$ sector. The field $\pi(x)$ is the Goldstone mode propagating at the speed of sound $c_s^2 = {1\over 2}$. The quadratic part of the action $S_\pi$ can be canonically quantized on the cylinder
\eqn\canquant{\eqalign{
\pi(\tau, {\bf n}) &= \pi_0 +  \pi_1 \tau +  \sum_{J>0,m}  \sqrt{4\pi \over \Omega_J} \left( a_{Jm} Y_{Jm} ({\bf n}) e^{-\Omega_J \tau} + a_{Jm}^\dagger Y_{Jm}^* ({\bf n}) e^{\Omega_J \tau} \right), \cr
\Omega_J &= \sqrt{J(J+1) \over 2} \ ,
}}
where $\pi_0, \pi_1$ are zero modes and canonical commutation relations are $[\pi_0, \pi_1] = 2$ and $[a_{Jm}, a_{J'm'}^\dagger] = \delta_{JJ'} \delta_{mm'}$.\foot{The Euclidian reality condition $\pi(-\tau,{\bf n})^\dagger = \pi(\tau, {\bf n})$ implies $\pi_0^\dagger = \pi_0, \pi_1^\dagger = -\pi_1$. Thus, we can write zero modes $\pi_0 = a^\dagger +a, \pi_1 = a^\dagger - a$ in terms of creation-annihilation operators with the latter acting on the vacuum in the standard way.} The free propagator
\eqn\propagator{
D(\tau,x) \equiv \la \pi(0, {\bf n}_2) \pi(\tau, {\bf n}_1) \ra \ ,
}
of $\pi$'s is given by the solution to Green's equation
\eqn\Green{
\left( \p_\tau^2 + {1\over 2} \triangle_{S^2} \right) D(\tau, x) = -4\delta(\tau) \delta(1-x) \ ,
}
where $x = {\bf n}_1 {\bf n}_2 = \cos \theta$ is the angle between two light operator insertions on $S^2$. Again, notice a peculiar ${1 \over 2}$ which is a consequence of conformal symmetry. The explicit form of the solution to \Green\ is given by
\eqn\piprop{
D(\tau,x) =  -|\tau| + \sum_{J=1}^\infty {2J+1 \over \Omega_J} e^{-\Omega_J |\tau|} P_J(x) \ .}
The expression \piprop\ suggests that  $D(\tau,x)$ is non-analytic at $\tau =0$. It is, however, manifest in \Green\ that $D(\tau,x)$ is analytic everywhere except at $\tau = 0$, $x = 1$ where two operators collide.

We would like to use EFT to compute correlation functions of light operators ${\cal O}_q$ in the background of the state created by the heavy operator. Any light operator ${\cal O}_q$ with scaling dimension $\Delta_q$ and charge $q$, both of order $O(1)$, can be represented at low energies in terms of Godstone boson degrees of freedom \MoninJMO
\eqn\lightopGB{
{\cal O}_q = c_q |\p \chi|^{\Delta_q} e^{i q \chi} + c_{q}^{{\cal R}} {\cal R} |\p \chi|^{\Delta_q-2} e^{i q \chi}  + \dots \ ,
}
where $c_q$ and $c_q^{\cal R}$ are constants not fixed by EFT and by ellipsis we denote further curvature couplings which lead to corrections suppressed at large $Q$. In practice, the expression for light operators \lightopGB\ should be expanded around the saddle \GBfluct
\eqn\lightopsaddle{\eqalign{
&{\cal O}_q (\tau, {\bf n}) = c_q \mu^{\Delta_q} e^{\mu q \tau} \times \cr
& ~~~~~ \times  \left( 1 + {iq\alpha \over 2\sqrt \mu} \pi(\tau, {\bf n}) +  {1\over 2} \left( iq\alpha \over 2 \sqrt{\mu}\right)^2 \pi^2(\tau, {\bf n}) +
{i\alpha \Delta_q \over 2\mu^{3/2}} \dot \pi (\tau, {\bf n}) + 2 {c_q^{\cal R} \over c_q} {1\over \mu^2} + ... \right) \ ,
}}
where we only kept terms which contribute to the correlators below at the order relevant for us. In \lightopsaddle\ $\pi^2(\tau, {\bf n})$ should be understood as a normal-ordered product.

Equations \saddleexp, \lightopsaddle\ provide us with a weakly coupled description of CFT in a state with large charge $Q$. Canonical quantization of the Goldstone $\pi$ gives the spectrum of operators in the charge $Q$ sector, as was found in \HellermanNRA. Further, using the representation of light operators \lightopGB, \lightopsaddle\ one can systematically compute correlators of the form
\eqn\EFTcorr{
\la {\cal O}_{Q} {\cal O}_{q_1} \dots {\cal O}_{q_n} {\cal O}_{-Q} \ra = \int D\chi \  {\cal O}_{q_1} \dots {\cal O}_{q_n} e^{-S_E} \ .
}
Now, we move on to describing the results of \HellermanNRA, \MoninJMO\ regarding the operator spectrum and correlation functions \EFTcorr.

%%%%%%%%%%%%%%%%%%%%%%%%%%%%%%%%%%%%%%%%%%%%%%%%%

\subsec{Operator Spectrum}
Using the operator/state correspondence one finds that the lowest dimension operator with large charge $Q$ has a scaling dimension \saddleexp
\eqn\DimLargeQ{
\Delta_Q = {2\over 3} \alpha  Q^{3/2} + \beta \sqrt{ Q } + C + O(Q^{-1/2}) \ .
}
The coefficients of the first two terms depend on the UV theory. On the other hand, the third term of order $O(1)$ is completely universal and given by $C = - 0.0937256\dots$. This is simply the Casimir energy of the Goldstone $\pi$.\foot{The value of $C$ was originally computed in {\HellermanNRA} and later corrected in {\MoninBWF}. }

The spectrum of low-lying operators is parametrized by integers $\vec{n} = (n_1, n_2, ...)$ and given by
\eqn\GBspectrum{
\Delta_Q^{\vec{n}}  = \Delta_Q + \sum_{J=1}^\infty n_J \Omega_J, \qquad \Omega_J  = \sqrt{J(J+1) \over 2} \ .
}
Each of the modes $\Omega_J$ corresponds to an excitation of the Goldstone boson $\pi$ with an angular momentum $J$ around the saddle {\saddle}. Excitations $\Omega_{J=1}$ are related to the descendants of primaries that appear in the $s$- and $u$-channel OPE. We will demonstrate this very explicitly shortly. Having $n_1$ modes $\Omega_{J=1}$ in \GBspectrum\ corresponds to the level $n_1$ descendant of $(0,n_2, n_3, \dots)$ with dimension $\Delta_Q^{(n_1,n_2,\dots)} = \Delta_Q^{(0,n_2,\dots)} + n_1$. The modes $\Omega_{J>1}$ correspond to new primary operators of various spins $j \leq \sum_{J=1}^{\infty} n_J J$.

Further, using the CCWZ prescription \lightopGB, the authors in \MoninJMO\ computed three- and four-point correlations functions. The results are as follows.

\subsec{Three-point Function}

We consider a three-point function of two heavy and one light operator
\eqn\threeCFT{\eqalign{
&\la {\cal O}_Q(x_{in}) {\cal O}_{-q}(x) {\cal O}_{-(Q-q)}(x_{out}) \ra =\cr
& ~~~~~~~ = { \lambda_{Q,-q, -(Q-q)} \over  |x_{out} - x_{in}|^{\Delta_{Q-q} + \Delta_Q - \Delta_q} |x_{out} - x|^{\Delta_{Q-q} - \Delta_Q + \Delta_q}|x - x_{in}|^{\Delta_Q - \Delta_{Q-q} + \Delta_q}} \ .
}}
To compute this three-point function in EFT, one has to slightly modify the path integral \EFTcorr\ to account for the extra charge $q$ in the final state. This is implemented by adding an extra term to the chemical potential $S_E \to S_E +i q \int {d {\bf n} \over 4\pi} \chi(\tau_{out}, {\bf n }) $. Using the prescription \GBfluct, \saddleexp, \lightopGB, \EFTcorr\ together with the mentioned modification, the EFT computation for the three-point function on the cylinder gives
\eqn\EFTthreepoint{\eqalign{
&\la {\cal O}_Q(\tau_{in}) {\cal O}_{-q}(\tau, {\bf n}) {\cal O}_{-(Q-q)}(\tau_{out}) \ra_{cyl} =
c_{-q} \mu^{\Delta_q }   e^{-\Delta_Q (\tau_{out} - \tau_{in})} e^{q\mu(\tau_{out} - \tau)} \times \cr
&\times   \Bigg( 1 + {(\alpha q)^2 \over 4\mu} \la \pi(\tau, {\bf n}) \int {d{\bf n}' \over 4\pi} \pi(\tau_{out},{\bf n}') \ra +  {(\alpha q)^4 \over 32\mu^2} \la \pi(\tau, {\bf n}) \int {d{\bf n}' \over 4\pi} \pi(\tau_{out},{\bf n}')\ra^2 - \cr
&- {\alpha^2 q \Delta_q \over 4\mu^2} \la \dot \pi(\tau, {\bf n}) \int {d{\bf n}' \over 4\pi} \pi(\tau_{out},{\bf n}') \ra  + {2c_q^{\cal R} \over \mu^2 c_q} + O(\mu^{-3}) \Bigg) \ ,
}}
where it is assumed that $\tau_{out} \to \infty , \tau_{in} \to - \infty$. Next, we can insert the expression for the propagator \piprop\ into \EFTthreepoint. The role of the integrals over ${\bf n}'$ is to project onto the zero mode in the propagator. Also changing large the $\mu$ expansion to a large $Q$ expansion via \saddle, we obtain
\eqn\EFTthreepointt{\eqalign{
&\la {\cal O}_Q(\tau_{in}) {\cal O}_{-q}(\tau, {\bf n}) {\cal O}_{-(Q-q)}(\tau_{out}) \ra_{cyl} = \cr
& ~~~~~=c_{-q} \alpha^{\Delta_q} Q^{\Delta_q/2 }  e^{-\Delta_Q (\tau_{out} - \tau_{in})}e^{q\left( \alpha\sqrt Q + {\beta \over 2 \sqrt Q} \right)(\tau_{out} - \tau)} \times \cr
& ~~~~~\times   \Bigg( 1   - \underbrace{{\alpha q^2 \over 4\sqrt Q} (\tau_{out} - \tau) +  {\alpha^2 q^4 \over 32 Q} (\tau_{out} - \tau)^2 }_{\rm{corrections\ to}\ \Delta_Q, \Delta_{Q-q}}
 -\underbrace{{ q \Delta_q \over 4 Q}   + {\beta \over 2\alpha} {\Delta_q \over Q} +{2c_{-q}^{\cal R} \over  c_{-q} \alpha^2 Q} }_{\rm{corrections\ to}\ \lambda_{Q,-q, -(Q-q)}} + O(Q^{-3/2}) \Bigg) \ .
}}
Using the map \mapop\ from the cylinder to the plane, one can check that \EFTthreepointt\ is a large $Q$ expansion of \threeCFT\ with $\lambda_{Q,-q, -(Q-q)}$ given by
\eqn\EFTthreelam{
\lambda_{Q,-q, -(Q-q)} = c_{-q} \alpha^{\Delta_q} Q^{\Delta_q /2} \left( 1 - {q \Delta_q \over 4Q} + {\beta \over 2\alpha} {\Delta_q \over Q} +  {2c_{-q}^{\cal R} \over \alpha^2 c_{-q} Q} + O(Q^{-3/2}) \right) \ .
}
In particular, notice the leading universal scaling $\lambda_{Q,-q, -(Q-q)} \sim Q^{\Delta_q/2}$, emphasized in \MoninJMO.

\subsec{Four-point Function}
In a similar fashion one can compute the four-point function \fourpoint, \gqdef\ of two heavy and two light operators\foot{Notice the term $\sim {1\over Q} D^2$ which is missing in the formula (8.20) in \MoninJMO.}
\eqn\EFTfourGB{\eqalign{
&g_q(z,\b{z}) = c_q c_{-q} \alpha^{2 \Delta_q}  Q^{\Delta_q} e^{- \alpha q \sqrt{Q} \tau} \Big( 1 - {\beta \over 2} {q \over  \sqrt{Q}} \tau + {\alpha \over 4}{ q^2 \over \sqrt{Q}}  D(\tau,x)     \cr
&+ {\beta \Delta_q \over \alpha Q } + {\beta^2 \over 8} {q^2 \over Q} \tau^2 - {\alpha \beta \over 8} {q^3 \over Q} \tau D(\tau,x)
-{ q \Delta_q \over 2Q} \p_\tau D(\tau,x) + {\alpha^2 \over 32} { q^4\over  Q} D(\tau,x)^2 + \cr
& ~~~~~~~~~~~~~~~~~~~~~ + {2 \over \alpha^2  Q} \left( {c_q^{\cal R} \over c_q} + {c_{-q}^{\cal R} \over c_{-q}}  \right) + O(Q^{-3/2})
 \Big) \ ,
}}
where $g_q(z,\b z)$ was defined in \gqdef\ and the relation between the cylinder $(\tau,x)$ and plane $(z,\b{z})$ coordinates is given in \map. The overall prefactor $Q^{\Delta_q} e^{- \alpha q \sqrt{Q} \tau} $ in \EFTfourGB\ comes from evaluating the two light operators \lightopGB\ on the saddle. The $D, \p_\tau D, D^2$ terms are quantum corrections to the leading answer. The remaining terms in \EFTfourGB, constant and terms with explicit $\tau$'s, come from using \saddle\ to convert $\mu$ into large $Q$ expansion, and the term in the third line comes from the curvature coupling in the light operator \lightopGB.

Let us discuss the structure of the formula \EFTfourGB\ in more detail. First, it is manifestly $s=u$ crossing symmetric \crossingequation. Changing $\tau \to -\tau$ , $q \to -q$ leaves invariant every term in \EFTfourGB\ (this can be seen using \piprop). Second, the result \EFTfourGB\ is analytic at non-coincident points, namely away from $\tau =0$, $x=1$. This follows from analyticity of the Goldstone propagator $D(\tau, x)$ defined by \Green. Third, due to conformal invariance and unitarity, it should be possible to decompose \EFTfourGB\ into a sum of conformal blocks with positive coefficients. Just from looking at \EFTfourGB\ it is not obvious that it is the case. Of course, it is guaranteed by the conformal symmetry of the action \EFTaction, but it will prove instructive to explicitly see how this happens. To avoid overwhelming the reader with too many equations, let us first discuss the four-point function to the order $O(Q^{-1/2})$.

\subsec{Four-point Function at ${1 \over \sqrt Q}$ Order}
At order $O(Q^{-1/2})$ the four-point function is given by the first line in \EFTfourGB. Using the propagator \piprop\ it can be cast into the form
\eqn\fourNLO{\eqalign{
g^{EFT}_q(z,\b{z})
&=c_q c_{-q}  \alpha^{2 \Delta_{q}} (z \b z)^{-{\alpha \over 2} q \sqrt{Q}} \Bigg( 1 + {1\over 4\sqrt{Q}} \left( -\beta q + {\alpha q^2 \over 2} \right) \log (z \b{z}) \cr
&+  {\alpha \over 4}{q^2 \over \sqrt{Q}} \sum_{J =1}^{\infty} {2J +1 \over \Omega_J } (z\b{z})^{{1\over 2}\Omega_J} P_J\left(x \right) +
O\left(Q^{-1} \right)
 \Bigg) \ ,
}  }
where the first line comes from the expansion of $(z\b{z})^{{1\over 2}(\Delta_{Q-q} - \Delta_Q) }$ at large $Q$ with $\Delta_Q, \Delta_{Q-q}$ being the dimensions of lightest operators in the sectors with charge $Q$ and $Q-q$ respectively, as given by \DimLargeQ.

The result \fourNLO\ was derived for $z \b{z} <1$ (equivalently $\tau<0$). To obtain the EFT answer for $z \b{z} >1$ (equivalently $\tau>0$) one simply needs to substitute $z \to {1\over z}$, $\b{z} \to {1\over \b{z}}$, $q \to -q$ in \fourNLO, namely the full correlator takes the form
\eqn\fullcorrelator{
g_q (z , \bar z) = \theta \left(1 - z \bar z \right) g^{EFT}_q(z, \bar z) +  \theta\left(z \bar z - 1\right) g^{EFT}_{-q} \left({1 \over z}, {1 \over \bar z} \right) \ .
}
The first term in \fullcorrelator\ gives the $s$-channel expansion for $z \b{z} <1$ and the second term gives the $u$-channel expansion for $z \b{z} >1$.\foot{This is somewhat reminiscent of the discussion in \HartmanOAA.} Indeed, in the $s$-channel formula \fourNLO\ the leading term is the contribution of the scalar with dimension $\Delta_{Q-q}$ and the term $J=1$ is the contribution of the first descendant of $\Delta_{Q-q}$, in accordance with the form of the conformal block \blockQ. In particular, no new primary operators with dimension $\Delta_{Q-q} + 1$ appear. The terms with $J \geq 2$ are primary operators with spin $J$ and dimensions $\Delta_{Q-q} + \Omega_J$.

In the form \fullcorrelator\ the four-point function is manifestly expanded into conformal blocks and trivially satisfies $s=u$ crossing $z \to {1\over z}$, $\b{z} \to {1\over \b{z}}$, $q \to -q$. On the other hand, the reader may be puzzled by an apparent non-analyticity of the formula \fullcorrelator\ at $z \b z = 1$. However, as we reviewed in the previous subsection, the correlator is analytic away from $z = \b z = 1$ and the only singularity is at $z = \b z = 1$ when two operators collide.  The $s$-channel OPE expansion in terms of EFT operators breaks down at $z \b z = 1$ and the $u$-channel expansion takes over at $|z|>1$.\foot{As shown in \PappadopuloJK\ the convergence of the $s$-channel OPE is optimal in terms of the so-called $\rho$-coordinate. The contribution of descendants, however, is not suppressed in the large $Q$ limit in the $\rho$-frame. This makes it unsuitable for the large $Q$ analysis.}

\vfil \break

\subsec{Four-point Function at ${1 \over Q}$ Order}

The main new feature at ${1 \over Q}$ order is the presence of an infinite number of operators of every spin in the OPE. Let us write down in detail the conformal block expansion of the correlator \EFTfourGB\
at the order $O(Q^{-1})$

\hfil\break
\eqn\gNNLO{\eqalign{
&g^{EFT}_q(z, \b{z}) = c_q c_{-q}  \alpha^{2 \Delta_{q}} (z \b z)^{-{\alpha \over 2} q \sqrt{Q}} \Bigg[ 1 + \cr
&+\underbrace{{1\over 4\sqrt{Q}} \left( -\beta q + {\alpha q^2 \over 2} \right) \log (z \b{z}) }_{\rm{correction\ to}\ \Delta_{Q-q}} +
\underbrace{{3\alpha \over 4} {q^2 \over \sqrt{Q}} (z \b{z})^{1\over 2} P_1}_{\rm{1st\ descendant\ of}\ \Delta_{Q-q}}
+\underbrace{{\alpha \over 4} {q^2 \over \sqrt{Q}} \sum_{J=2}^\infty {2J+1 \over \Omega_J} (z \b{z})^{{1\over 2} \Omega_J} P_J}_{\rm{primaries}\ \Delta_{Q-q}+\Omega_J} + \cr
&+ \underbrace{ {1\over 32 Q}  \left( -\beta q + {\alpha q^2 \over 2} \right)^2 \log^2 (z \b{z})}_{\rm{correction\ to}\ \Delta_{Q-q}}  +
\underbrace{{\alpha q^2\over 16 Q} \left( -\beta q + {\alpha q^2 \over 2} \right) \log (z \b{z}) \sum_{J=1}^\infty {2J+1 \over \Omega_J} (z \b{z})^{{1\over 2} \Omega_J} P_J}_{\rm{correction\ to}\ \Delta_{Q-q} + \Omega_J } - \cr
& - \underbrace{{3\over 2} {q \Delta_q \over Q} \sqrt{z \b{z}} P_1}_{\rm{1st\ descendant\ of}\ \Delta_{Q-q}} +
\underbrace{{\alpha^2 \over 32} {q^4 \over Q} (3 \sqrt{z \b{z}} P_1)^2 }_{\rm{2nd\ descendant\ of}\ \Delta_{Q-q}} +
\underbrace{{3\alpha^2\over 16} {q^4 \over Q}  \sum_{J=2}^\infty {2J+1 \over \Omega_J} (z \b{z})^{{1\over 2} (\Omega_J+1)} P_J P_1}_{\rm{1st\ descendant\ of}\  \Delta_{Q-q}+\Omega_J } - \cr
& -\underbrace{ {q \Delta_q \over 2 Q}}_{\delta \lambda_{Q,-q, -(Q-q)} } -
\underbrace{{q \Delta_q \over 2 Q} \sum_{J=2}^\infty (2J+1) (z \b{z})^{{1\over 2} \Omega_J} P_J}_{\delta \lambda_{Q,-q, -(Q-q)}^J}
+\underbrace{{\alpha^2 q^4 \over 32 Q} \left( \sum_{J=2}^\infty {2J+1 \over \Omega_J} (z \b{z})^{{1\over 2} \Omega_J} P_J \right)^2}_{\rm{primaries}\ \Delta_{Q-q} + \Omega_{J_1} + \Omega_{J_2}} + \cr
& +  \underbrace{{2 \over \alpha^2  Q} \left( {c_q^{\cal R} \over c_q} + {c_{-q}^{\cal R} \over c_{-q}}  \right)}_{\delta \lambda_{Q,-q, -(Q-q)}^J}
 \Bigg] \ ,
}}
where we indicated what is the interpretation of each term in $s$-channel conformal block expansion. Terms $\delta \lambda_{Q,-q, -(Q-q)}$ and $\delta \lambda_{Q,-q, -(Q-q)}^J$ stand for the corrections to the three-point functions of the $\Delta_{Q-q}$ and $\Delta_{Q-q} + \Omega_J$ correspondingly. The contribution of descendants in the fourth line of \gNNLO\ is in perfect agreement with the conformal blocks \blockQ, \leveltwoscalar.

The ${1 \over Q}$ terms in \gNNLO\ can be regrouped as in the second line in \EFTfourGB\ to make crossing manifest
\eqn\NNLO{\eqalign{
&{\beta^2 q^2 \over 32 Q} \log^2(z \b{z}) - {\alpha \beta q^3 \over 16 Q} \log(z\b{z}) \left( {1\over 2} \log(z\b{z}) +  \sum_{J=1}^\infty {2J+1 \over \Omega_J} (z \b{z})^{{1\over 2} \Omega_J} P_J \right) + \cr
&+ {\alpha^2 q^4 \over 32 Q} \Bigg[ {1\over 4} \log^2(z\b{z}) + \log(z\b{z})  \sum_{J=1}^\infty {2J+1 \over \Omega_J} (z \b{z})^{{1\over 2} \Omega_J} P_J +
(3\sqrt{z \b{z}} P_1)^2 + \cr
&+ 6  \sum_{J=2}^\infty {2J+1 \over \Omega_J} (z \b{z})^{{1\over 2} (\Omega_J+1)} P_J P_1 +{\alpha^2 q^4 \over 32 Q} \left( \sum_{J=2}^\infty {2J+1 \over \Omega_J} (z \b{z})^{{1\over 2} \Omega_J} P_J \right)^2 \Bigg] \cr
&- {q \Delta_q \over 2 Q} \sum_{J=0}^\infty (2J+1) (z \b{z})^{{1\over 2} \Omega_J} P_J
 +  {2 \over \alpha^2  Q} \left( {c_q^{\cal R} \over c_q} + {c_{-q}^{\cal R} \over c_{-q}}  \right).
}}
Thus, crossing invariant combinations are built from different types of corrections and there is a nontrivial interplay between them.

\subsec{Short Distance Limit and Regime of Validity}

It is instructive to write down a short distance expansion \MoninJMO\ of the four-point function \EFTfourGB\ in order to understand when the EFT approximation breaks down and to make connection with the macroscopic limit that we discuss in later sections. By short distances we mean the distance between the light operators becoming small. In terms of coordinates on the cylinder it corresponds to the region $\tau, \theta \to 0$. In the $(z, \b z)$ coordinates it corresponds to the region $z, \bar z \to 1$.

At small $\tau, \theta$ one can approximate $D(\tau,\theta)$ by a flat space propagator. Alternatively, one can compute it directly from \piprop. Indeed, the propagator $D(\tau,\theta)$ has a singularity at $\tau,\theta \to 0$, which is given by the large $J$ asymptotic of the sum \piprop\ $\sum_J e^{-{\tau \over \sqrt{2}} J} P_J$. This is the generating function for Legendre polynomials. Thus, we have
\eqn\propshort{
D(\tau,\theta) \approx {2\sqrt{2} \over \sqrt{ {1\over 2} \tau^2 + (\theta R)^2 } } \approx {4\sqrt{2} \over \sqrt{ {1\over 2} (2-z  - \b{z})^2 - (z - \b{z})^2  } }, \quad \tau, \theta \to 0, \quad z, \b{z} \to 1  \ .
}
Inserting this into \EFTfourGB, we have at small $\tau,\theta$
\eqn\fourshort{\eqalign{
g_q(z,\b{z}) = &c_0 Q^{\Delta_q} \Bigg( 1 + {\alpha \over \sqrt{2}} {q^2 \over \sqrt{Q}} {1\over \sqrt{ {1\over 2} \tau^2 + \theta^2 R^2 } }  + {\alpha^2 \over 4} {q^4 \over Q} {1\over \left( {1\over 2} \tau^2 + \theta^2 R^2 \right) } \cr
&+ {q \Delta_q \over \sqrt{2} Q} { \tau \over \left( {1\over 2} \tau^2 + \theta^2 R^2 \right)^{3/2} } + \dots
\Bigg) \ .
} }
We see that the large $Q$ expansion breaks down when $\tau, \theta \sim {1 \over \sqrt{Q}}$. In particular, the $t$-channel (light-light OPE) is not accessible within the EFT.

Let us now discuss in more details the regime of validity of EFT. It is supposed to be a good approximation when operator insertions are separated by distances much larger than the charge density scale. On the cylinder that means
\eqn\EFTregimeA{
(\tau_1 - \tau_2)^2 + \theta_{12}^2 R^2 \gg {R^2 \over Q} \ .
}

Similarly, the EFT breaks down close to the Lorentzian cone $\tau \to i t$ and $t_{12} = \theta_{12} R$. Indeed, in the light-cone limit the $s$-channel expansion (heavy-light OPE) is dominated by double-twist operators, which simply reproduce the identity operator in the $t$-channel (light-light OPE) \FitzpatrickYX, \KomargodskiEK. The whole EFT answer \fourshort\ gives a subleading contribution in the light-cone limit. What really breaks down in the light-cone limit is the matching \lightopGB\ of the light operators onto the Goldstone boson degrees of freedom. In the light-cone limit the high energy modes of ${\cal O}_q$ are excited and dominate the expansion \lightopGB.

An important lesson of this discussion is that there are contributions to the correlators which are not described by EFT and we need to make sure that they give a suppressed contribution in order for the EFT answer to be a reliable approximation. The leading EFT answer for the four-point function \EFTfourGB\ is $ \sim Q^{\Delta_q} e^{-\alpha q \sqrt{Q} \tau}$. So we have a situation of the type
\eqn\EFTregimeB{
G(z, \b{z}) = f_{\rm{non-EFT}}(z,\b{z}) + Q^{\Delta_q} e^{-\alpha q \sqrt{Q} \tau} f_{EFT}(z,\b{z}) \ ,
}
where $f_{\rm{non-EFT}}(z,\b{z})$ is the contribution of operators not described by EFT.\foot{For example, the identity operator, stress-tensor, $U(1)$ current in the $t$-channel or double-twist operators in the $s$-channel.} Thus, EFT is a good approximation to the correlator only when the second term in \EFTregimeB\ is exponentially large. Namely, EFT dominates the answer for the whole correlator at large $Q$ in \EFTregimeB\ when $q \tau <0$.

\subsec{Free Field Theories}

Let us contrast the results of EFT with correlators in free field theories. In the theory of a free complex scalar the two-point function takes the form
\eqn\twopoint{
\la \bar \phi (x) \phi(0)\ra = {1 \over |x|^{d-2}} \ .
}
For the heavy operator we choose ${\cal O}_{Q} ={1 \over (Q!)^{1/2}} \phi^Q$ and for the light operator we choose ${\cal O}_{-q} ={1 \over (q!)^{1/2}} \bar \phi^q$, where the normalization factors are such that the two-point functions are normalized to one.

The correlation function then takes the following form
\eqn\correlator{\eqalign{
&\la {\cal O}_Q(0) {\cal O}_{-q}(z,\bar z) {\cal O}_q(1) {\cal O}_{-Q}(\infty) \ra ={1 \over \left[ (1-z)(1-\bar z) \right]^{\Delta_{q}} } \times \cr
&\sum_{n=0}^{q} Q^n q^n {\prod_{k=0}^{n-1}(1-{k \over Q})(1-{k \over q}) \over (n!)^2}  \left( {(1-z)(1-\bar z)} \over z \bar z \right)^{n \Delta_{\phi}} ,
}}
where $\Delta_{\phi} = {d-2 \over 2}$ and for $n=0$ the product in the numerator is simply $1$. Note that this expression is only valid for $q>0$, so in a sense \correlator\ should be multiplied by $\theta(q)$. For $q<0$ the answer is different and cannot be obtained by changing $q \to - q$ in the expression above.

It is easy to take the large $Q$ limit of the correlator \correlator. We get
\eqn\largeQfree{
G_{q>0}(z,\bar z) = {Q^q \over  (z \bar z)^{q \Delta_{\phi}}} {1 \over \Gamma(q+1)}  \left(1 +  {q(1-q) \over 2 Q} + {q^2 \over Q} \left( {z \bar z \over (1-z)(1- \bar z)} \right)^{\Delta_{\phi}} + O\left({ Q^{-2}}\right)  \right) \ .
}
Note that we have the following identity
\eqn\identityB{
\left( {z \bar z \over (1-z)(1- \bar z)} \right)^{\Delta_{\phi}} =  \sum_{m=0}^{\infty} (z \bar z)^{\Delta_{\phi}+ {m \over 2}} C_m^{({d \over 2} - 1)} \left( {z + \bar z \over 2 \sqrt{z\bar z} } \right) \ .
}
The absence of the descendant in \largeQfree\ is due to the fact that in this case $\Delta_{Q-q} - \Delta_{Q,q} = 0$ (see \blockQ). Thus, we see that, on the  one hand, the structure of the correlator is very similar to the one appearing in the EFT. On the other hand, if we ignored $\theta(\pm q)$ and defined $g_q(z,\bar z)$ according to \gqdef\ and used \largeQfree\ for $G(z, \bar z)$, the result for $g_{q}(z,\bar z)$ would not be crossing symmetric. In the case of EFT $\theta(\pm q)$ did not arise.

For free charged fermions $\psi_{\alpha}, \b{\psi}_{\alpha}$ in $3d$ we have a heavy charged charged operator that corresponds to a state with a Fermi sea. It has a scaling dimension $\Delta_{Q} \sim Q^{3/2}$. The fluctuations around this state are, however, not described by a Goldstone boson EFT. In particular, correlators contain a non-analyticity in charges $\theta(\pm q)$, similar to the free complex boson discussed above.

%%%%%%%%%%%%%%%%%%%%%%%%%%%%%%%%%%%%%%%%%%%%%%%%%%%%%%%

\newsec{Macroscopic Limits of Correlators}
Now we switch gears and discuss a seemingly unrelated subject. However, we will soon use it as an important input to our bootstrap analysis in the next section.

A heavy operator with a large global charge corresponds to a state with large energy and charge densities on the cylinder ${\Bbb R} \times S^{d-1}$. The limit $Q \to \infty$ with $R$ fixed, where $R$ is the radius of $S^{d-1}$, results in both the energy and charge density going to infinity. In this context it is natural to consider a combined limit where both $Q \to \infty$ and $R \to \infty$ such that the correlation functions of light operators (that is, whose dimensions held fixed in the limit) in this state remain finite, up to an overall rescaling of the light operators.

Equivalently, as $Q \to \infty$, the background value of a light operator is given in terms of the three point function, $\lambda_{Q,-q, -(Q-q)}$, where $-q$ is the charge of the light operator and smoothness in $Q$ is assumed to relate the spectrum at charge $Q$ and $Q-q$. The macroscopic limit corresponds to the scaling regime of cross ratios in a four-point function with two light operators, such that the ratio of the four-point function and the product of the above three-point functions is kept fixed.

This happens when the two light operators are brought sufficiently close together. The result are correlation functions of the flat space CFT in a nontrivial state. In various situations, the energy and/or charge density may remain fixed or be scaled to 0 in this limit. We call these type of limits {\it macroscopic}. It will turn out to be useful for solving bootstrap equations in the next section.

The limit $R \to \infty$ when the energy density is kept fixed, also known as the thermodynamic limit, was recently discussed in the CFT context in \LashkariVGJ. We will consider this limit in our case as well, even though the correlators that we will get in this limit are not thermal. Rather, they are described by the Goldstone EFT. More generally, the limit with fixed energy density does not have to coincide with the macroscopic limit (in which correlators are kept finite) for every state, even though it is not clear to us what are all cases in which it fails. Not very surprisingly, they fail to agree in the case of a free complex scalar field, as we will review. It is also easy to show that it will not exist for BPS operators with $\Delta \sim Q$ and, more generally, we expect it to fail for chiral ring operators in CFTs with moduli. In this case the macroscopic limit has rather different properties, as we discuss below.

The basic idea is the following. Consider a CFT on the cylinder ${\Bbb R} \times S^{d-1}$
\eqn\metric{\eqalign{
d s^2_{cyl} = d \tau^2 + R^2 d \Omega_{d-1}^2 &= \left( {R \over r} \right)^2 \left( d r^2 + r^2 d \Omega_{d-1}^2 \right) = \left( {R \over r} \right)^2 d s^2_{R^d}, ~~~ \tau = R \log r \  .
}}
The mapping of correlators to the plane takes the following form
\eqn\mapping{\eqalign{
\la H | O(x) ... | H \ra_{cyl} &= \left( {r \over R} \right)^{ \Delta_O} {\la O_{H}(0) O(x) ... O_{H}^{\dagger} (\infty)  \ra_{{\Bbb R}^d}  \over \la O_{H}(0) O_{H}^{\dagger} (\infty) \ra_{{\Bbb R}^d}}\ ,
}}
where we suppressed other operators and their conformal transformation factors. A heavy operator insertion corresponds to a state on the cylinder with energy $E = {\Delta_H \over R}$ and, in general, some charge $Q$.

We can consider the limit $R \to \infty$, simultaneously with $\Delta_{H} \to \infty$ and $Q \to \infty$, such that
the energy density $\eps$ and the charge density $q$
\eqn\energyandchargedensity{
\eps = {\Delta_H \over R^{d}}, ~~~ q = {Q \over R^{d-1}} \ .
}
are kept fixed and non-zero. For this to be possible the scaling dimension $\Delta_H$ and the charge $Q$ should be related
as follows
\eqn\relation{
\Delta_{H} \sim Q^{{d \over d-1}} \ ,
}
as should be clear from \energyandchargedensity. We, thus, first discuss the case \relation\ and later consider the most general possibility.

In what follows, it will be useful for us to study correlation functions in the macroscopic limit. To be more specific let us consider the following $(n+3)$-point function on ${\Bbb R}^{d}$
\eqn\nplustwo{\eqalign{
G(z_i ,\bar z_i) &\equiv \la O_{H}(0) O_{L_1} (z_1, \bar z_1) \ ... \  O_{L_n} (z_n, \bar z_n) O_{L_{n+1}} (1) O_{H}^{\dagger}(\infty) \ra \ ,
}}
where for simplicity we put all the operators in one plane. The most generic case is analogous.

We would like to describe the macroscopic limit above in the conformal invariant terms. Taking the radius of the sphere $R$ to infinity and keeping the distance between the light operators $L$ fixed is conformally equivalent to keeping the sphere intact and scale local operators toward each other. This becomes a limit in the space of cross ratios.

In terms of cross-ratios $z_i ,\b z_i$ \nplustwo\ the physical distance between light operators is $L \sim R|1-z_i|$, which corresponds to taking the limit
\eqn\thermodynamic{
z_i = 1 - {w_i \over  \Delta_{H}^{{1/d}} }, ~~~ \bar z_i = 1 - {\bar w_i \over  \Delta_{H}^{{1/d}}  } \ .
}
where we take $\Delta_H \to \infty$ and $w_i, \b{w}_i$ - fixed. In writing \thermodynamic\ we used that $R \sim \Delta_{H}^{{1 \over d}} \to \infty$, which follows from \energyandchargedensity .

The statement that the macroscopic limit exists for the correlation functions, thus, becomes
\eqn\macroonecross{
G^{\eps}(w_i , \bar w_i) \equiv \lim_{\Delta_H \to \infty} \Delta_H^{- {1\over d} \sum_{i = 1}^{n+1} \Delta_{L_i}  \ } G\left(1 - {w_i \over \Delta_{H}^{{1 / d}} }, 1 - {\bar w_i \over \Delta_{H}^{{1/ d}} }\right) \ .
}
The pre-factor $R^{- \sum_i \Delta_{L_i}}$ in \macroonecross\ is due to the conformal factor in \mapping.

Let us apply \macroonecross\ to the case of one light operator, which corresponds to $n=0$. In this case the dependence on $(w_i, \bar w_i)$ trivializes and we get
\eqn\onepoint{
G^{\eps} = \lim_{\Delta_H \to \infty} \Delta_H^{- {\Delta_{L} \over d} } c_{HH^\dagger L} \ .
}
The existence of the macroscopic limit, thus, immediately implies that \LashkariVGJ \foot{See comments after equation $(9)$ in that paper.}
\eqn\boundonthreep{
c_{HH^\dagger L} \leq c_0 \Delta_H^{ {\Delta_{L} \over d} } \ ,
}
for some constant $c_0$ and any light operator $O_{L}$. Only operators that saturate the bound \boundonthreep\ contribute to the macroscopic limit. Both the identity operator and the stress tensor saturate the bound \boundonthreep. Indeed, for $T_{\mu \nu}$ we have $\Delta_{T_{\mu \nu}} = d$ and $c_{HH^\dagger T_{\mu \nu}} \sim \Delta_{H}$ so that \boundonthreep\ is saturated. For the conserved current we have $c_{HH^\dagger J_{\mu}} \sim Q$ and it saturates the bound \boundonthreep\ when \relation\ holds.

Further, one can use the bound \boundonthreep\ to consider a contribution of light operators into the $(n+3)$-point function \macroonecross\ in the light-light OPE channel. Each light operator $L'$ contributes in the macroscopic limit \thermodynamic\ $\sim ([1-z][1-\bar z])^{{\Delta_{L'} \over 2}} c_{H H^\dagger L'} \sim  \Delta_H^{ -{ {\Delta_{L'} \over d}} } c_{H H^\dagger L'}  $. Thus, only operators that saturate the bound \boundonthreep\ contribute in the finite energy density limit. The presence of identity operator and stress-tensor implies that the $(n+3)$-point function $ G^\e  (w, \bar w)$ is nontrivial, if exists.

For us the relevant example is $n=1$, namely the case of the four-point function. We have for the macroscopic limit \macroonecross
\eqn\macroonecross{
G^{\eps}(w , \bar w) \equiv \lim_{\Delta_H \to \infty} \Delta_H^{- {2\over d}  \Delta_{L}  \ } G\left(1 - {w \over \Delta_{H}^{{1 / d}} }, 1 - {\bar w \over \Delta_{H}^{{1/ d}} } \right) \ .
}
It is instructive to take the t-channel OPE expansion \blockst\ and see how each separate conformal block contributes in the macroscopic limit. We get
\eqn\macrotOPE{
G^{\eps}(w , \bar w) = (w \bar w)^{- \Delta_{L}} \sum_{ {\cal O}_{\Delta, J} } \lambda_{L,L, {\cal O}_{\Delta, J}} \lambda_{H,H , {\cal O}_{\Delta, J}} g_{\Delta, J}^{0,0} \left( {w \over \Delta_{H}^{{1 / d}} },{\bar w \over \Delta_{H}^{{1 / d}} } \right) \ .
}
Only primary operators that saturate the bound \boundonthreep\ give a non-zero contribution to \macrotOPE. Using \confmblock\ it is also easy to understand that descendants decouple, since extra powers of $(1-z)$ are suppressed by ${1 \over \Delta_{H}^{{1 / d}}}$. The contribution of light operators in the $t$-channel, thus, takes the form
\eqn\macrotOPEb{\eqalign{
G^{\eps}(w , \bar w) &= (w \bar w)^{- \Delta_{L}} \sum_{ {\cal O}_{\Delta, J} } \lambda_{L,L, {\cal O}_{\Delta, J}} \tilde \lambda_{{\cal O}_{\Delta, J}}  (w \bar w)^{{\Delta \over 2}} C_j^{({d \over 2}-1)} \left(w + \b{w} \over 2\sqrt{w\b{w}} \right) + ... \ , \cr
\tilde \lambda_{{\cal O}_{\Delta, J}} &= \lim_{\Delta_H \to \infty} \lambda_{H,H , {\cal O}_{\Delta, J}} \Delta_H^{- {\Delta \over d} } \ ,
}}
where the sum is over primary operators $ {\cal O}_{\Delta, J} $ saturating the bound \boundonthreep\ and by ellipsis we denoted a potential contribution of operators whose dimensions scale with $\Delta_H$ as well.
Note that we do not know the convergence properties of the $t$-channel OPE after taking the macroscopic limit. At the very least it should be a reliable asymptotic series for small $w \bar w$.

\subsec{Macroscopic Limit in the EFT}

In EFT the relation \relation\ holds and, therefore, we can take the limit $R \to \infty$ with finite energy and charge densities described above. In three dimensions $d=3$ the limit \thermodynamic\ becomes
\eqn\thermolimitEFT{
z = 1 - {w \over \sqrt{Q}}, \qquad \b{z} = 1 - {\b w \over \sqrt{Q}} \ .
}
On the cylinder \map\ in coordinates $\tau, \theta$ the limit \thermolimitEFT\ is equivalent to\foot{Recall that we are working in euclidian kinematics, where $w, \b w$ are complex conjugates of each other.}
\eqn\thermoEFTcylinder{
\tau \approx - {{\rm Re}\ w \over \sqrt{Q}}, \qquad \theta \approx {{\rm Im}\ w \over \sqrt{Q}} \ ,
}
up to corrections suppressed at large $Q$. To derive the macroscopic limit of the EFT four-point function \EFTfourGB, we simply insert \thermoEFTcylinder\ into the short-distance expansion \fourshort\ and obtain
\eqn\EFTmacro{\eqalign{
G^\e(w, \b w) &= \lim_{Q \to \infty} Q^{-\Delta_q} G \left(  1 - {w \over \sqrt{Q}},  1 - {\b w \over \sqrt{Q}} \right) = \cr
&= c_0 Q^{\Delta_q} e^{{\alpha \over 2} q (w + \b w) } \Bigg( 1 +  { \alpha q^2 \sqrt{2} \over \sqrt{ {1\over 2} (w + \b w)^2 + (w - \b w)^2  } } + \cr
& ~~~~~ +  {\alpha^2 q^4 \over  {1\over 2} (w + \b w)^2 + (w - \b w)^2    } -
 {2 \sqrt{2} q \Delta_q (w + \b w) \over \left[ {1\over 2} (w + \b w)^2 + (w - \b w)^2   \right]^{3/2} } + \dots
\Bigg) \ .
} }
In the macroscopic limit \thermolimitEFT\ the EFT regime of validity \EFTregimeA\ becomes $|w| \gg 1$. In this regime the expansion \EFTmacro\ is a controlled approximation with corrections suppressed by inverse powers of $w, \b w$.

Note, that when we take the macroscopic limit the structure is slightly different from the large $Q$ limit, namely the contributions which were parametrically suppressed in the large $Q$ limit could become of the same order in the macroscopic limit. This is essentially due to the fact that $z^{c \alpha(Q)} = (1 - {w \over \alpha(Q)})^{c \alpha(Q)} \to e^{- c w}$ when $\alpha (Q) \to \infty$. Therefore, for operators with different $c$, say $c_1$ and $c_2$, the one with larger $c$ is exponentially suppressed with respect to the one with smaller $c$ in the large $Q$ limit. While in the macroscopic limit they both become of the same order. Still, operators with larger $c$ stay exponentially suppressed in limit $|w|\gg 1$.

\subsec{Other Limits}

We can also imagine a situation when \relation\ does not hold. A well-known example of this type is $\Delta(Q) \sim Q$, which is common in supersymmetric and free theories. In this case it is clear that the limit we described above does not exist. Indeed, the conserved current $J_{\mu}$ would violate \boundonthreep
\eqn\violation{
c_{HH^\dagger J_\mu} \sim  Q > c_0 \Delta_H^{{\Delta_J \over d}} \sim Q^{{d-1 \over d}} \ .
}

More generally, if we imagine $\Delta(Q) \sim Q^{\alpha}$ then for $\alpha \geq {d \over d-1}$ the bound \boundonthreep\ is satisfied by the current and the thermodynamic limit described above might exist, whereas for $\alpha < {d \over d-1}$ the bound \boundonthreep\ is violated due to \violation.

In situations when the thermodynamic limit does not exist we could imagine a different limit
\eqn\macroonecrossAlpha{
G^{\beta}(w_i , \bar w_i) \equiv \lim_{\Delta_H \to \infty} \Delta_H^{- \beta \sum_{i = 1}^{n+1} \Delta_{L_i}  \ } G\left(1 - {w_i \over \Delta_H^{\beta}} , 1 - {\bar w_i \over \Delta_H^{\beta}} \right)
}
for some $\beta$. The condition that such a limit exists implies that
\eqn\boundonthreepAlpha{
c_{HH^\dagger L} \leq c_0 Q^{ \beta {\Delta_{L} } } \ .
}
For large enough $\beta$ we expect that the limit exists and is trivial. The question then is what is $\beta$ for which the limit exists and is nontrivial. This is controlled by the operator, whose three-point function first saturates the bound  \boundonthreepAlpha.

The macroscopic limit of the type \macroonecrossAlpha\ exists in the free complex scalar theory, as we describe below, and more generally we expect that it is relevant for CFTs with moduli spaces.

\subsec{Macroscopic Limit for the Free Complex Scalar}
Let us again contrast EFT with the theory of a free complex scalar, discussed in section 4.6. In this case we have $\Delta_Q \sim Q$. The proper macroscopic limit in this case is a zero energy and charge density limit of the type \macroonecrossAlpha\ with $\beta = {1 \over d-2}$. The operator that first saturates the bound \boundonthreepAlpha\ is the scalar $\bar \phi \phi$. The result for the correlator \correlator\ $\la \phi^Q \bar \phi^q \phi^q \bar \phi^Q \ra$ in this limit is given by
\eqn\macro{
G(w, \bar w) = \lim_{Q \to \infty} Q^{-q} G\left(1 - {w \over Q^{{1 /( d-2)}} }, 1 - {\bar w \over Q^{{1 / ( d-2 )}} }\right) = {L_q(- (w \bar w)^{(d-2)/2}) \over (w \bar w)^{q(d-2)/2}} \ ,
}
where $L_n(x)$ is the Laguerre polynomial. In this limit both the energy and the charge density go to zero. Nevertheless, the limit is nontrivial. A simple computation shows that \macro\ coincides with the two-point function on the moduli space $\la O_{-q}(w, \bar w) O_q(0)\ra_{\la \phi \ra = 1}$. Thus, in this case the macroscopic limit describes correlators on the moduli space.

%%%%%%%%%%%%%%%%%%%%%%%%%%%%%%%%%%%%%%%%%%%%%%%%%%%%%%%%%%%%%%%%%%%%%%%%

\newsec{Bootstrap at Large $Q$}

In the previous section we reviewed two particular solutions to the large $Q$ crossing: EFT and free theories. In this section we explore the structure of a general solution based on unitarity, crossing and the structure of the macroscopic limit.

We assume throughout that all operators that enter the large $Q$ crossing belong to the families which have a smooth dependence on $Q$.

\subsec{Crossing For The Vacuum}

As we discussed in section 2, in the large $Q$ limit at fixed $z$ we expect the operator with minimal dimension to dominate. In the EFT, the leading contribution to the correlator was given by a single scalar operator which belonged to the same family $\Delta_Q$ as the external state. In $d$ dimensions the large $Q$ asymptotic of the dimension is
\eqn\dimlargeQ{
\Delta_Q = {d-1\over d} \alpha Q^{d/(d-1)} + ... \ .
}
One might consider other possibilities as well. For instance, supersymmetric theories with BPS operators have $\Delta_Q \sim Q$. Here we focus on the case \dimlargeQ, which is expected to hold in generic interacting CFTs. Moreover, we will focus on $3$ dimensions, generalizing only some of the formulae to arbitrary $d$.

The crossing equation with a single scalar operator \dimlargeQ\ that dominates takes the form
\eqn\sucrossing{\eqalign{
g_{q}(z,\bar z) &= |\lambda_{Q,-q, -(Q-q)}|^2 (z\b{z})^{{1\over 2} (\Delta_{Q-q} - \Delta_Q )}  + ... \cr
&= |\lambda_{Q,q, -(Q+q)}|^2 \left( {1 \over z\b{z} } \right)^{{1\over 2} (\Delta_{Q+q} - \Delta_Q   )} + ... = g_{-q} \left({1 \over z}, {1 \over \bar z} \right) \ ,
}}
where we kept only the leading term of the conformal block \blockQsub. This equality implies that to leading order we have
\eqn\identitycross{
\Delta_{Q-q} - \Delta_Q = \Delta_Q - \Delta_{Q+q}  \ .
}
This is indeed true if all three operators belong to the same family $\Delta_Q$ \dimlargeQ. Then \identitycross\ becomes a continuity equation for $- q {\p \Delta_Q \over \p Q}$ as a function of $Q$.

Similarly, \sucrossing\ implies equality of the corresponding three-point functions to leading order
\eqn\qeventhreepoint{
|\lambda_{Q,-q, -(Q-q)}|^2 = |\lambda_{Q,q, -(Q+q)}|^2  \ .
}

A simple consequence of this matching is that we cannot add a finite number of operators to \sucrossing\ without spoiling crossing. Indeed, imagine that in  \sucrossing we had instead
\eqn\correction{
g_{q} (z, \bar z) =  |\lambda_{Q,-q, -(Q-q)}|^2 (z\b{z})^{{1\over 2} (\Delta_{Q-q} - \Delta_Q )} \left( 1 + \sum_{i=1}^{N} (z \bar z)^{\delta_i \over 2} C_{J_i}^{({d \over 2}-1)} \left(z + \b{z} \over 2\sqrt{z\b{z}} \right) \right) .
}
If the operators that enter the $s$- and $u$-channels depend on charge smoothly, then $\delta_{i}$ are independent of $q$. As a result crossing \sucrossing\ implies that
\eqn\deltazero{
\delta_i = 0 \ .
}
In other words, the vacuum could be degenerate and could carry spin, but all the excitations should be suppressed by ${1 \over Q}$.

Another possibility is that $N = \infty$ in \correction. This case lies beyond the scope of the present work.

Assuming that macroscopic finite energy density limit \macroonecross\ exists, we immediately find using the leading answer for the four-point function \sucrossing\ that
\eqn\bound{
\lim_{Q \to \infty} |\lambda_{Q,-q, -(Q-q)}|^2 \Delta_Q^{-{2 \over d}\Delta_q} = {\rm const} \ .
}
This is indeed the case in the EFT as can be seen from \EFTthreelam\ after setting $d=3$
\eqn\threepointscal{
\lambda_{Q,-q, -(Q-q)} \sim \Delta_Q^{\Delta_q/d} \sim Q^{\Delta_q /(d-1)} \ .
}

\subsec{Crossing at Subleading Order }

The strategy that we adopt is to approach $|z|=1$ both from the $s$-channel and the $u$-channel and make sure that they match smoothly. The leading order correction appearing at order $1\over \sqrt{Q}$ takes the form (compare with \fourNLO)
\eqn\NLO{\eqalign{
f(\tau, \theta) &= - | \tau | + 3 e^{-|\tau|} P_1(x) + \sum_{J=0}^{\infty} c_J e^{- \e_{J} | \tau |} P_{J}( x), \quad x= \cos \theta, ~~ c_J >0 \ ,
}}
where the first term in $f$ is the correction to the scaling dimension of ${\cal O}_{Q-q}$ and the second term is the first descendant. These two terms are necessarily present due to the leading order scalar ${\cal O}_{Q-q}$. The sum over $J$ represents new primary operators appearing at this order. Analyticity at $|z|=1$ which is the same as analyticity at $\tau = 0$ is not manifest due to the non-analyticity of $|\tau| = \theta(\tau) \tau - \theta(-\tau) \tau$. If we formally compute $\p^n_\tau f(0,\theta)$, there will be terms that involve $\delta(\tau)$ and its derivatives. For the function to be analytic away from the point $(\tau, \theta) = (0,0)$ where light operators collide, these terms should be set to zero. This condition leads to the following set of equations (independent smoothness conditions involve only odd derivatives of $f$)
\eqn\conditions{
\delta_{0,n} + 3P_1(x)  + \sum_{J=0}^{\infty} c_J \eps_J^{2n + 1} P_{J}(x)  =0 \ , ~~~ x = \cos \theta \neq 1, ~~~ n = 0,1,2,... \ ,
}
where the sum should be understood as a limit of the regulated expression \NLO\ (alternatively, we can use any other regulator, see below).

In the case of the EFT $\e_J = \sqrt{J(J+1) \over 2}, c_J = {2J+1 \over \e_J}, J \geq 2$ one can easily check that \conditions\ is satisfied using the generating functional for Legendre polynomials
\eqn\genfunc{
{1 \over \sqrt{1 - 2 x t + t^2}}  = \sum_{J=0}^\infty  P_J(x) t^J \ ,
}
by application of a proper combination of $t \p_t$ derivatives at $t=1$. The singularity that appears on the RHS of \conditions\ at $x=1$ in EFT case is a linear combination of $\delta^{(n)}(1-x), \delta^{(n-1)}(1-x), \dots$, as one can simply check by recalling the identity
\eqn\identitydelta{
{1 \over 2} \sum_{J=0}^{\infty} (2 J + 1) P_J(x) P_J(y) = \delta(x-y)  \ .
}
and its derivatives at $y=1$.

We would like to understand if the EFT solution is the unique solution to \conditions\ given the form of the correction \NLO. To do that it is useful to better understand the microscopic origin of $\delta(1-x)$ in \identitydelta. As we will see, the precise form of $c_J, \e_J$ is related to the behavior of $f(\tau, \theta)$ close to $\tau=0$, $x=1$.\foot{Note, that this singularity has nothing to do with very short-distance $t$-channel regime, which we effectively collapsed to a point when taking $Q \to \infty$ limit. Rather, it is coming from the macroscopic limit region, as we will indeed see later.} For now we assumed that we have one operator of each spin in \NLO\ and will consider generalizations later.

\ifig\zplaneintegral{The cross ratio $z$-plane. The region $|z|<1$ is described by the $s$-channel OPE in the large $Q$ limit, whereas the region $|z|>1$ is described by the $u$-channel OPE. We consider the integral of the derivatives of the correlation function along the drawn contours, namely we take the difference between the derivative evaluated slightly outside and slightly inside the circle. The smoothness of the correlator away from $z=1$ implies that the difference in that region is of $O(\eps)$. On the other hand, we can use the OPE to see that the result should be $O(1)$. This means that the integral is controlled by the singularity close to $z = 1$. In this region the correlator is governed by its properties in the macroscopic limit. The shaded area denotes the short distance region relevant for this computation.} {\epsfxsize2.5in\epsfbox{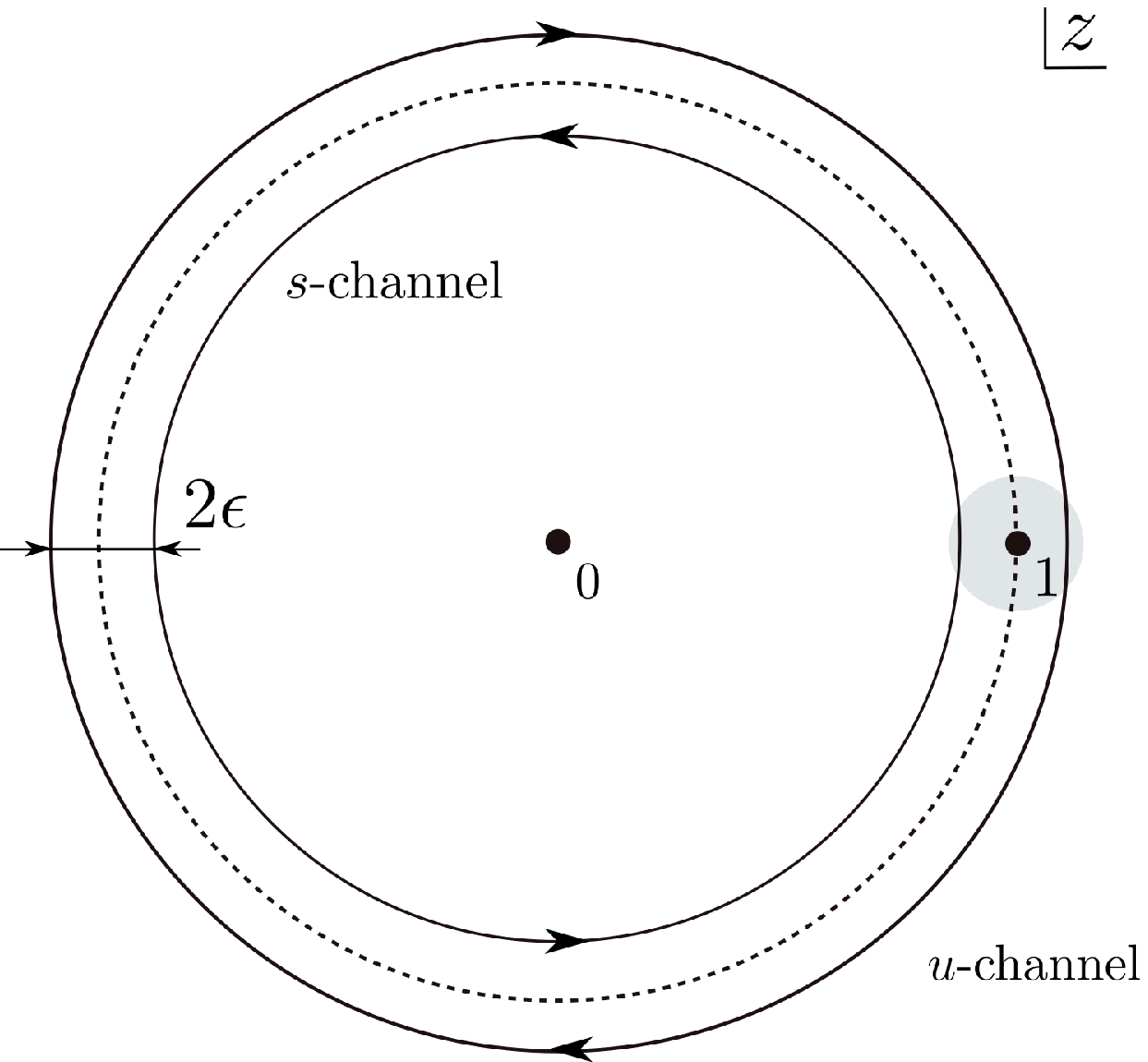}}

The strategy  we adopt is illustrated in \zplaneintegral. We consider the discontinuities of $\tau$-derivatives of \NLO\ at $\tau = \e \to 0$ and integrate both sides with a Legendre polynomial.  We find
\eqn\ce{\eqalign{
&\delta_{n,0}\delta_{J,0} + 3 \delta_{J,1} +   c_J \e_J^{2n+1} = \cr
& ~~~~~ -{1\over 4} (2J+1) \lim_{\e\to 0} \int_0^{\pi}   d\theta \sin \theta  \left[  \p_\tau^{2n+1} f(\e,  \theta) - \p_\tau^{2n+1} f(-\e,  \theta) \right]  P_J(\cos \theta) \ .
} }
Note that due to the smoothness of $f(\tau,  \theta)$ away from $\tau = \theta = 0$ most of the integral produces a contribution of $O(\eps)$, whereas the LHS of \ce\ is $O(1)$. The sum rule \ce, thus, makes it manifest that $c_J, \e_J$ are controlled by the behavior of $f(\tau, \theta)$ near the singularity.\foot{In principle, all the steps that we perform here could be repeated for a generic CFT. In this case instead of one operator of each spin on the LHS of \ce\ we have an infinite number of them. This structure makes the result much less powerful, see \PappadopuloJK. The benefit of the large $Q$ limit is to make the number of Regge trajectories, that appear on the LHS of \ce, finite.}

The macroscopic limit discussed in section 5 provides us with a controllable short-distance approximation of the correlator and of the function $f(\tau,\theta)$. We have
\eqn\expansion{
G(z, \bar z) \sim 1 + {1 \over \sqrt{Q} } f(\tau,  \theta) \ .
}
Assuming that the macroscopic limit exists, we conclude that $f(\tau,  \theta)$ can at most grow as $\sqrt Q$ in this limit
\eqn\fmacro{
f\left( \tau = - {{\rm Re}\ w \over \sqrt{Q}}, \theta = { {\rm Im}\ w \over \sqrt{Q}} \right) \sim \sqrt{Q} \ .
}
If the growth is slower than $\sqrt{Q}$, there will be no solution for $c_J, \e_J$, as will become clear shortly. So it must grow exactly as $\sqrt{Q}$. This immediately implies that the leading term in $f(\tau,\theta)$ at small distances must be a homogeneous function of degree one
\eqn\fscaling{
f(\lambda \tau, \lambda \theta) \approx  \lambda^{-1} f(\tau, \theta) , \quad \tau, \theta \ll 1 \ .
}
The most general solution of this homogeneity equation is
\eqn\fsolhom{
f(\tau,\theta)  = {1\over \sqrt{\tau^2 + \theta^2 }} F(y) + \dots, \quad y = {\tau \over \sqrt{\tau^2 + \theta^2 }}, \quad \tau,\theta \ll 1 \ ,
}
where $F(y)$ is an arbitrary function regular on an interval $y \in [0,1]$ and $...$ stands for less singular terms.\foot{In fact, it is regular in a larger domain, but this interval is all we need.} Indeed, the end points $y = 0$ and $y = 1$ correspond to $\tau = 0, \theta \neq 0$ and $\tau \neq 0 ,\theta = 0$ respectively. The correlator is regular at these points and we can smoothly interpolate between them.  Moreover, the function $F(y)$ is even $F(y) = F(-y)$ due to crossing.
In the small $\tau$ and $\theta$ limit the two cross-ratios become $u \approx 1 + 2\tau, v \approx \tau^2 + \theta^2$.

Let us first consider $n=0$ in  \ce . To evaluate the integral we can simply insert the expression \fsolhom\ into \ce.
Any less singular term in \fsolhom\ which we denoted by $+...$ produces a contribution which is vanishing in the $\eps \to 0$ limit. To reproduce the LHS of \ce\ we can, therefore, use the short distance approximation $f(\tau,\theta)  = {1\over \sqrt{\tau^2 + \theta^2 }} F(y) $. Plugging \fsolhom\ into \ce\ we find for $n=0$
\eqn\cesol{\eqalign{
\delta_{J,0} + 3 \delta_{J,1} + c_J \e_J &= {1\over 2} (2J+1)\int_0^1   dy   \left( F(y)  + y F'(y) - {1\over y} F'(y) \right) ,
}}
where we expanded
\eqn\expansion{
\sin \theta \ P_{J} (\cos \theta) = \sum_{m=0}^{\infty} \alpha_m \theta^{2 m + 1} \ ,
}
 switched to an integral over $y = {\e \over \sqrt{\e^2 + \theta^2}}$, and took the limit $\e \to 0$. The known coefficients $\alpha_m$ are polynomials in $\Omega_J^2 = {J(J+1) \over 2}$ of the maximal power $\alpha_m \sim ( \Omega_{J}^{2})^m$. For $n=0$ only the leading term $m=0$ contributes.

The integral in \cesol\ is finite since $F(y)$ is a regular even function on the interval $[0,1]$. In particular, the last term is finite since $F'(y) \sim y$ near $y=0$. Finally, we obtain
\eqn\cesolution{
\delta_{J,0} + 3 \delta_{J,1} + c_J \e_J = \beta_0^{(0)} ( 2J+1 ) \ ,
}
where $\beta_0^{(0)}$ is an unknown $J$-independent constant given by the integral in \cesol.

Next, we consider the case of $n=1$ in \ce\ for regularity of $\p_\tau^3 f$. There are two changes compared to the case of $n=0$. First, the terms with $m=0$ and $m=1$ in the expansion \expansion\ contribute. Second, some of the less singular terms which we have not written explicitly in \fsolhom\ could generate a finite contribution as well. Let us present the final result and then make a few comments
\eqn\ceee{\eqalign{
3\delta_{J,1} + c_J \e_J^3 &= - {1\over 4} (2J+1) \lim_{\e\to 0} \int_0^{\pi}   d\theta \sin \theta  \left[  \p_\tau^3 f(\e,\theta) - \p_\tau^3 f(-\e, \theta) \right] P_J\left( \cos \theta \right) = \cr
& = (2J+1)(\beta_0^{(1)}  + \beta_1^{(1)} \Omega_J^2), \qquad
\Omega_J^2 \equiv {J(J+1) \over 2} \ .
} }

The limit $\e \to 0$ is somewhat subtler in this case. After inserting $ {1\over \sqrt{\tau^2 + \theta^2 }} F(y) $ in the first line of \ceee\ and changing integration variable to $y = {\e \over \sqrt{\e^2 + \theta^2}}$, one finds the leading term of order $1\over \e^2$. It, however, should vanish if we are to obtain finite answer in the limit $\e \to 0$. Indeed, one can check that the $y$-integral multiplying $1\over \e^2$ is zero for any $F(y)$. The finite contribution comes from two terms. The term $\beta_1^{(1)}$ arises solely from the leading short distance asymptotic in \fsolhom\ multiplying the second term $m=1$ in \expansion. The term that involves  $\beta_0^{(1)}$, on the other hand, receives a contribution from a less singular term in \fsolhom\ multiplying $m=0$ term in \expansion. The coefficients $\beta_0^{(1)}, \beta_1^{(1)}$ are related to the precise form of the function $F(y)$ and the subleading terms in \fsolhom\ and cannot be fixed. The dependence on $J$, however, is completely fixed in \ceee.

Similar manipulations give equations for any $n = 0,1,2,\dots$ in \ce. The result is simply
\eqn\smooth{
\delta_{n,0} \delta_{J,0} + 3\delta_{J,1} + c_J \e_J^{2n+1} = (2J+1) W_{n}(\Omega_J^2), \qquad n = 0,1, 2, \dots
 \ , }
where
\eqn\polynomial{
W_{n}(\Omega_J^2) = \sum_{k=0}^{n} \beta_{k}^{(n)} \Omega_J^{2k}
}
is a polynomial of degree $n$ whose coefficients are arbitrary and related to the precise form of the function $F(y)$ and less singular terms in \fsolhom. The self-consistency of \ce\ requires that all the terms singular in $\eps$ integrate to zero. As in example above, the maximal power of the polynomial $W_n(\Omega_J^2)$ is controlled by the leading asymptotic of \fsolhom.\foot{The structure of the equations is reminiscent of the Casimir trick \AldayEYA\ used in the conformal bootstrap. }

\subsec{The Smoothness Conditions}

Next, we would like to analyze in more details the equations \smooth. So far we assumed that there is exactly one operator of each spin \NLO. Let us start by relaxing this condition. We can generalize the ansatz \NLO\ to have $N$ Regge trajectories, i.e. $N$ operators with each spin with squares of three-point functions and scaling dimensions $c_{J,i}, \e_{J,i}$ and $i = 1,\dots, N$. On general grounds we expect that the OPE data $c_{J,i}, \e_{J,i}$ is analytic in spin $J$ for some $J \geq J_0$. As was shown in \Caron, $J_0 = 2$. For $J=0,1$ the physical values of the OPE coefficients $(c_{i j k}^{phys}, \Delta_{J}^{phys})$ and the values that one obtains from the analytic expressions $(c_{i j k}^{analytic}, \Delta_{J}^{analytic})$ do not have to agree.

In this way natural generalizations of \NLO\ take the following form
\eqn\newansatz{\eqalign{
f(\tau, \theta) &= - |\tau| + 3 x e^{- |\tau|} + \sum_{J=2}^\infty \sum_{i=1}^{N} {2 J + 1 \over \eps_{J,i}} d_{J,i} e^{- \eps_{J,i} |\tau|} P_J(x) \cr
&+\sum_{i=1}^{N_0} {1 \over E_{0,i}} D_{0,i} e^{- E_{0,i} |\tau|} + 3 x \sum_{i=1}^{N_1}  {1 \over E_{1,i}} D_{1,i} e^{- E_{1,i} |\tau|} \ ,
}}
where we parameterized the squares of three-point functions by $d_{J,i}, D_{0,i}, D_{1,i}$. This ansatz consists of $N$ Regge trajectories together with a finite number $N_0$ of scalar operators and $N_1$ spin one operators. As discussed above, $\eps_{J,i}$ and $d_{J,i}$ are analytic functions of $J$ for ${\rm Re}[J] \geq 2$. Unitarity implies that the squares of three-point functions are positive and scaling dimensions are real. Moreover, we assume that scaling dimensions $\e_{J,i},E_{0,i}, E_{0,1}$ are also positive\foot{This positivity is not necessary and one can consider cases when this does not hold. However, we restrict our analysis to positive $\e_{J,i},E_{0,i}, E_{0,1}$. This assumption is equivalent to clustering in the large $Q$ limit that we discussed in section 2.}
\eqn\unitaritythree{
d_{J,i}, D_{1,i}, D_{0,i}, \e_{J,i}, E_{0,i}, E_{0,1}>0 \ .
}
The case when some of the three-point functions or energies are zero reduces to the one with smaller $N$, $N_0$, $N_1$. We also assume that
\eqn\nonequal{
\eps_{J,i} \neq \eps_{J,j} , ~~~E_{0,i} \neq E_{0,j},~~~ E_{1,i} \neq E_{1,j} 
}
for $i \neq j$, since otherwise the solution is again equivalent to one with smaller $N$, $N_0$, $N_1$.

Let us first analyze the analytic part of \newansatz\ which is encoded in $\eps_{J,i}$ and $d_{J,i}$. The corresponding generalization of equations \smooth\ takes the form
\eqn\analyticequations{
\sum_{i=1}^{N} d_{J,i} \eps_{J,i}^{2 n} = W_{n} (\Omega_{J}^2) \ , ~~~ n=0,1,2, \dots, \quad J \geq 2\ .
}
Originally, these equations are written for integer spins $J=2,3,4,...$. However, analyticity in $J$ of both the LHS and the RHS implies that \analyticequations\ should hold in the whole $J$ plane, due to the Carlson theorem.

For $J=0,1$ the smoothness conditions take the form
\eqn\smoothnesslowspin{\eqalign{
J=0&: ~~~ \delta_{n,0}  + \sum_{i=1}^{N_0} D_{0,i} E_{0,i}^{2n} = W_n(0), ~~~ n = 0,1, ... \ , \cr
J=1&: ~~~ 1  + \sum_{i=1}^{N_1} D_{1,i} E_{1,i}^{2n} = W_n(1), ~~~ n = 0,1, ... \ .
}}
Equations \analyticequations, \smoothnesslowspin\  comprise the full set of smoothness conditions of the function \newansatz. Let us describe their solutions. We start with the smoothness conditions \analyticequations\ for $J \geq 2$. Then we consider equations \smoothnesslowspin\ and show that EFT is the unique solution for one Regge trajectory $N=1$.

\subsec{Solution of Smoothness Conditions for $J \geq 2$.}

Let us introduce the notation\foot{This $z$ parameterizes dependence on spin and has nothing to do with $z$ in previous sections, which was the position of an operator insertion.}
\eqn\notation{
z \equiv \Omega_J^2, \quad d_i(z) \equiv d_{J,i}, \quad \e_i(z) \equiv \e_{J,i}^2 \ ,
}
where analyticity in spin for ${\rm Re}\ J \geq 2$ implies that the functions $d_i(z), \e_i(z)$ are analytic for ${\rm Re}\ z \geq \sqrt{3}$. Moreover, for real $z \geq \sqrt{3}$ the functions $d_i(z), \e_i(z)$ are real and positive and $\e_i(z) \neq \e_j(z)$ for $i\neq j$. Then the equations \analyticequations\ take the form
\eqn\analyticeq{
\sum_{i=1}^N d_i(z) \e_i(z)^n = W_n(z), \quad n = 0,1,2,\dots \ .
}
The solutions of these equations are derived in appendix A. The result is as follows. The scaling dimensions $\e_i(z)$ are given by the $N$ distinct solutions of the $N$-th order algebraic equation
\eqn\epssolpol{
\prod_{i=1}^N (x - \e_i(z)) = x^N - {\cal P}_1(z) x^{N-1} + {\cal P}_2(z) x^{N-2} + \dots + (-1)^N {\cal P}_N(z) =0 \ ,
}
where ${\cal P}_n(z)$ is an $n$-th order polynomial in $z$.\foot{These solutions can be traced back to the 17th century works of Albert Girard and Isaac Newton \Newton.}
 The three-point functions $d_i(z)$ are given by
\eqn\dsol{
d_k(z) =  \prod_{i<j} [\eps_j(z) - \eps_i(z)]^{-1} \det \left(  \matrix{
      1 & \dots & W_0 & \dots & 1 \cr
      \e_1(z) & \dots & W_1(z) & \dots & \e_N(z) \cr
        & \dots  & \dots & \dots  &  \cr
        \e_1(z)^{N-1} & \dots & W_{N-1}(z) & \dots & \e_N(z)^{N-1}
        } \right) \ ,
}
where the $k$-th column is made of polynomials $W_n(z)$. The equations \epssolpol, \dsol\ give a complete solution of the smoothness conditions \analyticeq\ for spins $J \geq 2$. Finally, one needs to impose constraints on $W_n(z), {\cal P}_i(z)$ to ensure that $\e_i, d_i(z)$ are real and positive for real $z\geq \sqrt{3}$. One simple consequence of this is that ${\cal P}_i(z)$ is real and positive for real $z\geq \sqrt{3}$.

The solution \epssolpol, \dsol\ is parameterized by $N(N+2)$ real constants contained in polynomials $W_0 , \dots, W_{N-1}, {\cal P}_1, \dots ,{\cal P}_N$.

Now, let us explain why \epssolpol, \dsol\ is a solution. The equation \dsol\ is simply the solution of the first $N$ equations in \analyticeq\ considered as a linear system for $d_i(z)$ due to Cramer's rule. The solution for $\e_i(z)$ comes from the rest of the equations in \analyticeq. Generically, the scaling dimensions $\e_i(z)$, being the solutions of the equation \epssolpol, will contain branch cuts and will define an analytic function on an $N$-sheeted Riemann surface. Then the LHS of \analyticeq\ sums over all sheets of this Riemann surface and cancels the branch cuts to reproduce the polynomial on the RHS.

Further, $d_k(z)$ \dsol\ has poles at $\e_k = \e_j, j\neq k$. This pole is cancelled on the LHS of \analyticeq\ between $d_k$ and $d_j$, which is guaranteed by \dsol. Essentially, this follows from the anti-symmetry of the determinant in \dsol\ under exchange of two columns.

Finally, the choice of powers of polynomials ${\cal P}_n(z)$ ensures correct behavior $d_i(z) \sim 1, \e_i(z) \sim z $ when $z \to \infty$ to match the behavior of the RHS of \analyticeq.

The proof of the uniqueness of the solution \epssolpol, \dsol\ and more details about its derivation can be found in appendix A.

\subsec{One Regge Trajectory}

Now, we can consider the smoothness conditions \smoothnesslowspin\ for $J = 0,1$. Unlike the situation for $J \geq 2$, these equations do not have a nice analytic structure. Therefore, we are forced to deal with them separately for different numbers of Regge trajectories. We start with $N = 1$.

In this case \epssolpol, \dsol\ are reduced to
\eqn\oneregge{
\e_J^2 = {\cal P}_1(\Omega_J^2) =  c^2 \Omega_J^2 +m^2, \qquad  d_J = W_0, \qquad J \geq 2 \ ,
}
where $c,m, W_0$ are arbitrary parameters. The polynomials $W_n$ are given by
\eqn\Woneregge{
W_n(\Omega_J^2) = W_0 (c^2 \Omega_J^2 + m^2 )^n, \quad n = 0,1,2,\dots \ .
}
Then, we would like to solve the equations \smoothnesslowspin\ for $J=0,1$. It is convenient to consider two cases separately:

1) $\underline{N_0 = 0}$. The equation \smoothnesslowspin\ for $J=0$ gives
\eqn\jzerosol{
W_0 = 1, \quad m = 0 \ .
}
For spin $J=1$ we have
\eqn\jonesol{\eqalign{
&1 + \sum_{i=1}^{N_1} D_{1,i}  =1 \ , \cr
&1 + \sum_{i=1}^{N_1} D_{1,i} E_{1,i}^2  = c^2 \ ,  \cr
&1 + \sum_{i=1}^{N_1} D_{1,i} E_{1,i}^4  = c^4 \ . \cr
&\dots
}}
Since $D_{1,i} >0$, the only solution is $N_1 = 0, c =1$. All free parameters are fixed in this case and we recover the EFT solution
\eqn\EFTsol{
f = - |\tau| + 3 x e^{-|\tau|} + \sum_{J=2}^\infty  {2J+1 \over \Omega_{J}}   e^{-\Omega_{J} |\tau|} P_J(x) \ .
}

2) $\underline{N_0 > 0}$. In this case \smoothnesslowspin\ for $J=0$ gives
\eqn\jzerosoll{\eqalign{
&1 + \sum_{i=1}^{N_0} D_{0,i}  =W_0 \ , \cr
&\sum_{i=1}^{N_0} D_{0,i} E_{0,i}^{2n}  = W_0 m^{2n},\quad n = 1,2,\dots \ .
}}
These do not have nontrivial unitary solutions. Indeed, without loss of generality we assume that $E_{0,1}^2 > \dots > E_{0,N_0}^2$. Taking the large $n$ limit of the second equation in \jzerosoll\ we find $D_{0,1} E_{0,1}^{2n} = W_0 m^{2n}, n \gg 1$. This implies $D_{0,1} = W_0, E_{0,1} = m, N_0 = 1$, which contradicts the first equation in \jzerosoll.

We, thus, conclude that the EFT of \HellermanNRA, \MoninJMO\ is the unique solution to the crossing equations if we assume that only one Regge trajectory appears in the OPE.

\subsec{Two Regge Trajectories}

For two Regge trajectories the equations \epssolpol, \dsol\ reduce to
\eqn\twoRegge{\eqalign{
&x^2 - {\cal P}_1(\Omega_J^2) x + {\cal P}_2(\Omega_J^2) = 0 \ , \cr
&d_1 = {W_0 \e_2 - W_1 \over \e_2 - \e_1} , \quad d_2 = {W_1 - W_0 \e_1 \over \e_2 - \e_1}, \quad J \geq 2 \ ,
}}
where the scaling dimensions $\e_i$ are the solutions of the first equation in \twoRegge\ and the dependence on $\Omega_J^2$ is suppressed for brevity. Thus, we have
\eqn\twoReggee{\eqalign{
&\e_1(\Omega_J^2) = {1\over 2} \left( {\cal P}_1 + \sqrt{ {\cal P}_1^2 - 4 {\cal P}_2} \right), \quad \e_2(\Omega_J^2) = {1\over 2} \left( {\cal P}_1 - \sqrt{ {\cal P}_1^2 - 4 {\cal P}_2} \right) \ , \cr
&d_1(\Omega_J^2) = {W_0 \over 2} + { W_1 - {1 \over 2} W_0 {\cal P}_1 \over \sqrt{ {\cal P}_1^2 - 4 {\cal P}_2} }, \quad
d_2(\Omega_J^2) = {W_0 \over 2} - { W_1 - {1 \over 2} W_0 {\cal P}_1 \over \sqrt{ {\cal P}_1^2 - 4 {\cal P}_2} } , \quad J \geq 2 \ .
}}
Now we need to solve the smoothness conditions \smoothnesslowspin\ for $J = 0,1$. Instead of writing down the most general solution, we will consider two simple examples, which will demonstrate what kind of solution one might have. The first solution will have one Regge trajectory with a dispersion relation of a free massive particle and one Regge trajectory of the Goldstone mode. The other solution will have neither a Goldstone mode nor any other interpretation in terms of quasiparticles and might be considered as some strongly interacting CFT.

As the first example let us consider the following solution
\eqn\twoReggeone{\eqalign{
&f(\tau,\theta) = D(\tau,x) + (W_0 - 1) D_{c, m}(\tau, x)  = \cr
& = - |\tau| + \sum_{J=1}^\infty {2J+1 \over \Omega_J} e^{-\Omega_J |\tau|} P_J(x) +
(W_0 - 1)\sum_{J=0}^\infty {2J+1 \over \sqrt{c^2 \Omega_J^2 + m^2}} e^{-\sqrt{c^2 \Omega_J^2 + m^2} |\tau|} P_J(x)
}}
where $D(\tau,x)$ is the propagator of the Goldstone boson \Green, \piprop\ and $D_{c,m}(\tau,x)$ stands for a propagator of a free particle of mass $m$ moving at the speed $c\over \sqrt{2}$, satisfying the equation
\eqn\massiveprop{
(\p_\tau^2 + {c^2 \over 2} \triangle_{S^2} - m^2) D_{c,m}(\tau,x) = -4 \delta(\tau) \delta(x-1) \ .
}
One can check that \twoReggeone\ is indeed a particular case of \twoReggee\ and that it solves the $J = 0,1$ constraints \smoothnesslowspin\ as well.

As the second example, let us consider \twoReggee\ and set $W_1 = {1\over 2} W_0 {\cal P}_1$ to cancel the second term in $d_i$. To further simplify our lives, let us take $N_0 = N_1 = 0$ in \smoothnesslowspin. Now it is easy to solve \smoothnesslowspin\ and find that $W_0 = 1, W_1(z) = z, {\cal P}_1(z) = 2z, {\cal P}_2(z) = (1-a^2)z^2 + a^2 z$ and $d_1 = d_2 = {1\over 2}, \e_{1,2} = z \pm a z \sqrt{z -1}$ with an arbitrary constant $a$. Thus we have
\eqn\twoReggenonpert{\eqalign{
&f(\tau,\theta) = - |\tau| + 3e^{-|\tau|}x + {1\over 2} \sum_{J=2}^\infty {2J+1 \over   \sqrt{ \Omega_J^2 + a \Omega_J \sqrt{\Omega_J^2 - 1}} }
e^{-|\tau| \sqrt{ \Omega_J^2 + a \Omega_J \sqrt{\Omega_J^2 - 1}}} P_J(x) + \cr
& + {1\over 2} \sum_{J=2}^\infty {2J+1 \over   \sqrt{ \Omega_J^2 - a \Omega_J \sqrt{\Omega_J^2 - 1}} }
e^{-|\tau| \sqrt{ \Omega_J^2 - a \Omega_J \sqrt{\Omega_J^2 - 1}}} P_J(x) \ .
}}
For $a=0$ we recover EFT, but for $a \neq 0$ this solution does not have a quasiparticle interpretation.

%%%%%%%%%%%%%%%%%%%%%%%%%%%%%%%%%%%%%%%%%%%%%%%%%%%%%%%%%%%%%

\newsec{Extensions}

In this section we sketch several extensions of the previous analysis. First, we consider external operators that carry spin. Second, we comment on correlation functions at the next order in ${1 \over Q}$. Third, we consider the correlation function in the limit $w \gg \bar w \gg 1$ and discuss matching to the $t$-channel OPE expansion.

\subsec{External Operators With Spin}

In the sections above the external operator was a heavy charged scalar. It corresponds to the Goldstone vacuum. On the other hand, if we act on the vacuum with the Goldstone creation operators $a^{\dagger}_{J,m} | Q \ra$ we end up with the state which corresponds to a primary operator $O^{\Delta_Q + \Omega_J , J}_{\mu_1 ... \mu_J}$ of spin $J$ and dimension $\Delta_Q + \Omega_J$. Taking this operator as an external state, we can repeat the computation of both the three- and four-point functions. We limit ourselves only to the leading nontrivial corrections in this case.

The three-point function takes the form
\eqn\threepointspin{\eqalign{
&\la Q-q, J, m' |  {\cal O}_{-q}(\tau, {\bf n}) |Q, J , m \ra_{cyl} = \cr
& = c_q \mu^{\Delta_q} e^{-\mu q \tau} \left( \la J, m' |  J , m \ra_{cyl} - {1 \over 8} {q^2 \alpha \over \sqrt{Q}}  \la J, m' |  \pi^2(\tau, {\bf n}) | J , m \ra_{cyl} \right) \cr
&=c_q \alpha^{\Delta_q} Q^{{\Delta_q \over 2}} e^{- \alpha \sqrt{Q} q \tau} \left( \delta_{m,m'} \left(1 - {\beta \over 2} {q \over  \sqrt{Q}} \tau\right) - {1 \over 4} {q^2 \alpha \over \sqrt{Q}} {4 \pi \over \Omega_J} Y_{J m}({\bf n}) Y^*_{J m'}({\bf n}) \right) ,
}}
where $|Q, J, m \ra = a^{\dagger}_{J,m}|Q \ra$ and we used \canquant\ to compute the second line. One can relate this result to a more familiar basis of structures in flat space considered in \CostaMG. For example, $\delta_{m,m'}$ corresponds to $H_{13}^J$, whereas $Y_{J m}({\bf n}) Y^*_{J m'}({\bf n})$ involves $V_1^J V_3^J$.

Next, let us compute the four-point function. The result takes the form
\eqn\object{\eqalign{
&\la Q,J, m' | O_{q}(\tau, {\bf n}_1 ) O_{-q}(0, {\bf n}_2) |Q, J , m \ra =c_q c_{-q} \alpha^{2 \Delta_q} Q^{\Delta_q} e^{- \alpha \sqrt{Q} q \tau} \cr
&\Big( \delta_{m,m'} \left(1 - {\beta \over 2} {q \over  \sqrt{Q}} \tau + {\alpha \over 4}{ q^2 \over \sqrt{Q}}  D(\tau,x) \right)- \cr
&- {1 \over 4}{\alpha q^2 \over \sqrt{Q}} {4 \pi \over \Omega_J} \left[ Y_{J,m}({\bf n}_1) - Y_{J,m}({\bf n}_2)  ] [Y_{J,m'}({\bf n}_1) - Y_{J,m'}({\bf n}_2) \right] \Big) \ ,
}}
where the only difference with the scalar case is the possibility of contracting the Goldstone field with the external state. This is the source of the spherical harmonics $Y_{J,m}({ \bf n})$ in the third line of \object. Note that the second and the third lines are crossing symmetric by themselves. Moreover, the second line, that is the $\delta_{m,m'}$ term, is identical to the scalar correlator. Decomposing this expression into conformal blocks we find the tower of operators with dimensions $\Delta_H + \Omega_J + \Omega_{J'}$.

Again, in principle, we could have used the technology of \CostaDW\ to decompose this expression into conformal blocks. This is, however, completely unnecessary, since by inserting a complete set of states in \object\ we explicitly get a sum over exchanges of states $| Q-q, J, \tilde m \ra$ as well as $|Q-q, J ,  m_1 ; J',  m_2 \ra$ with positive coefficients. The only nontrivial terms which are not of this type are $- {\beta \over 2} {q \over  \sqrt{Q}} \tau , {\alpha q^2 \over 4 \sqrt Q} \tau $ coming from the correction to the dimension of $\Delta_Q$, and $J=1$ term in the expansion of the propagator $D(\tau, x)$ which corresponds to the contribution of the descendant.

Let us check that the descendant comes with the correct coefficient. The first level descendants contribute as follows
\eqn\descednantcontribution{\eqalign{
&\sum_{\mu} \sum_{m'} {\la J, m | O_{q}(\tau, {\bf n}_1 ) P_{\mu} | J , m' \ra \la J , m' | K_{\mu}  O_{-q}(0, {\bf n}_2) | J,m \ra \over \la J , m' | K^{\mu} P_{\mu} | J , m' \ra } \cr
&=- \sum_{\mu} \sum_{m'} {\la J, m | [P_{\mu}, O_{q}(\tau, {\bf n}_1 ) ] | J , m' \ra \la J , m' | [K^{\mu} ,  O_{-q}(0, {\bf n}_2)] | J,m \ra \over \la J , m' | [K^{\mu} , P_{\mu}] | J , m' \ra } \cr
&=\alpha^2 q^2 Q {{\bf n}_1 . {\bf n}_2 \over 2 {2 \alpha \over 3} Q^{{3/2}}} c_q c_{-q} \alpha^{2 \Delta_q} Q^{\Delta_q} e^{- \alpha \sqrt{Q} q \tau} e^{\tau} \cr
&=c_q c_{-q} \alpha^{2 \Delta_q} Q^{\Delta_q} e^{- \alpha \sqrt{Q} q \tau} \left( {3 \over 4} {\alpha q^2 \over \sqrt Q} x e^{\tau} \right) \ ,
}}
which is exactly what we have in \object . In evaluating \descednantcontribution\ we used that $[K_{\mu}, P_{\nu}]=2 D \delta_{\mu \nu} - 2 M_{\mu \nu}$, as well as the action of $P_{\mu}$ and $K_{\mu}$ on the primaries which can be found, for example, in \PappadopuloJK\ (see formula (3.11) in that paper).

In principle, at this point we can repeat the bootstrap analysis. The only complication is the fact that external operators in this case carry spin and multiple tensor structures have to be taken into account. We leave the detailed analysis of this case for the future, but it is clear that the same type of structure appears as with the external scalar operator. Namely, requiring smoothness of the correlator would lead to a similar set of equations. Assuming that only one Regge trajectory is present, the solution is given by the Goldstone propagator. This time, however, we have operators of the type $\Delta_H + \Omega_J + \Omega_{J'}$. Then we can make this ``two-particle'' state an external operator and repeat the argument. In this way, we see that the spectrum of operators exhibits an additive structure \GBspectrum\ , as observed in the EFT.

\subsec{Comment on Bootstrap at Order ${1 \over Q}$}

In our analysis of the correlator it was crucial that the number of Regge trajectories that appear at leading order is finite. One can wonder if this structure persists at higher orders in ${1 \over \sqrt Q}$.

Consider the correction to the correlator of the type
\eqn\correction{
\delta G(z, \bar z) \sim {1 \over Q} \hat f(\tau,  \theta) \ .
}
From the existence of the macroscopic limit we learn that $\lambda^2 \tilde f(\lambda \tau, \lambda \theta)$ is finite in the limit of small $\lambda$. This means that when we compute the contribution of the spin $J$ operators at order ${1 \over Q}$, by an argument identical to the one in section 6.2, we potentially get a ${1 \over \eps}$ singularity
\eqn\ceNNLO{\eqalign{
&-{1 \over 2 J + 1}\lim_{\e \to 0} \sum_i c_{J,i} \eps_{J,i} e^{- \eps_{J,i} \eps}=\cr
&=\lim_{\e\to 0} \int_0^{\pi}   d\theta \sin \theta  \left[  \p_\tau \hat f(\e, \theta) - \p_\tau \hat f(-\e, \theta) \right]  P_J(\cos \theta) \sim {1 \over \eps} \ .
}}
If the coefficient in front of ${1 \over \eps}$ happens to be zero the situation is identical to the one encountered at leading order. Otherwise, \ceNNLO\ is consistent if the number of operators with spin $J$ on the LHS of \ceNNLO\ is infinite. This situation is identical to the discussion of Tauberian theorem in \PappadopuloJK . This is also precisely what happens in the case of EFT.

At this point a careful reader could be puzzled by how little mileage we get from the constraints in this case compared to the leading correction, where it was possible to bootstrap the answer completely. The crucial point is that in this case we have an infinite number of Regge trajectories on the LHS of \ceNNLO. A remarkable efficiency of this simple matching at the order $1\over \sqrt Q$ was due to a finite number of Regge trajectories. Here we see that at the order $1\over Q$ the equation \ceNNLO\ requires that
\eqn\ceNNLOb{\eqalign{
{1 \over 2 J + 1}\sum_i c_{J,i} \eps_{J,i} e^{- \eps_{J,i} \eps} = {2 \over \eps} \int_0^1   dy\   y \left( 2 \hat F(y)  + y \hat F'(y) - {1\over y} \hat F'(y) \right) , \quad \e \to 0 \ ,
}}
where we used $ \hat f(\tau, \theta) = {1 \over \tau^2 + \theta^2} \hat F({\tau \over \sqrt{\tau^2 + \theta^2}})$. We have not explored if \ceNNLOb\ being zero is consistent with the leading order solutions found in the previous sections.

\subsec{``Light-cone'' Bootstrap in the Macroscopic limit: $w \gg \bar w \gg 1$}

Let us recall the structure of the macroscopic limit in the $t$-channel. The macroscopic limit of the $t$-channel conformal block is given by $Q^{- {\Delta \over 2}} (w \bar w)^{{\Delta \over  2}} P_{J}({w + \bar w \over 2 \sqrt{ w \bar{w}}})$, where descendants are again further suppressed. We do not know the convergent properties of the $t$-channel OPE after we take the macroscopic limit. It should be definitely a reliable expansion for $w \bar w \ll 1$, but in this section we assume that it converges for $w \bar w \gg 1$ as well.

We can try to match the macroscopic limit of the EFT result \EFTmacro\ to the $t$-channel. The EFT description is only valid for $w, \b w \gg 1$, therefore, we get the following relation
\eqn\relationMacroT{\eqalign{
c_0 e^{{\alpha \over 2} q (w + \b w) } \left( 1 + O\left({1 \over w}, {1 \over \bar w}\right) \right) &= \sum_{\Delta, J} Q^{- {\Delta \over 2}}c_{\Delta, J}(w \bar w)^{{\Delta - 2 \Delta_q \over  2}} P_{J} \left({w + \bar w \over 2 \sqrt w \sqrt{\bar{w}}} \right) , ~~~ w, \bar w \gg 1 \ , \cr
c_{\Delta, J} &= \lambda_{Q,-Q, O_{\Delta, J}} \lambda_{q,-q, O_{\Delta, J}} \ , ~ \
}}
where for simplicity we kept only the leading order answer on the LHS. It is clear from \relationMacroT\ that only operators that saturate the bound on the three-point function \boundonthreep\ $\lambda_{Q,-Q, O_{\Delta, J}} \sim Q^{{\Delta \over 2}}$ contribute in the macroscopic limit. We can integrate over $x = {w + \bar w \over 2 \sqrt{ w \bar{w}}}$ to project both sides on the sector with given spin and derive an asymptotic density of states in each spin sector that is dictated by the LHS in \relationMacroT.

In the limit $w \gg \bar w \gg 1$ the mapping can be made more explicit. In this case the argument of the Legendre polynomial is $x \gg 1$ and the leading asymptotic of the block simply becomes
\eqn\asymptotic{
c_0 e^{{\alpha \over 2} q w } \left(1 + O \left({\bar w \over w}\right)  \right)= \sum_{\Delta, J} {4^{-J}  \Gamma(1+2 J) \over \Gamma(1+J)^2} Q^{- {\Delta \over 2}} c_{\Delta, J}  (w \bar w)^{{\Delta - J - 2 \Delta_q \over  2}} w^J , ~~~ w \gg \bar w \gg 1 \ .
}
There are natural candidate operators on the RHS to reproduce the LHS. These are the usual double-twist operators ${\cal O}_{q} \pa^J {\cal O}_{-q}$ which have an asymptotic twist $\Delta - J = 2 \Delta_q$. Note, however, the difference with the usual light-cone bootstrap. Here, the spin $J$ is not the largest parameter in the problem and we first take the large $Q$ limit. Remembering that $c_{\Delta, J} = \lambda_{q, -q, O_{q} \pa^J O_{-q}} \lambda_{Q, -Q, O_{q} \pa^J O_{-q}}$ and using the usual light-cone bootstrap result  \FitzpatrickYX, \KomargodskiEK\ for $\lambda_{q, - q, O_{q} \pa^J O_{-q}}$
\eqn\lightconeThree{
\lambda_{q, - q, O_{q} \pa^J O_{-q}} = {\pi^{1 \over 4} J^{\Delta_q - {3 \over 4}} \over 2^{J + \Delta_q - 1} \Gamma(\Delta_q)} + ...
}
we can derive a formula for $\lambda_{Q, -Q, O_{q} \pa^J O_{-q}}$. In this way we get
\eqn\asymptotic{
c_0 e^{{\alpha \over 2} q w }= \sum_{J} {1 \over J^{1/2} \pi^{1/2}} {\pi^{1 \over 4} J^{\Delta_q - {3 \over 4}} \over 2^{J + \Delta_q - 1} \Gamma(\Delta_q) } Q^{- {2 \Delta_q + J \over 2}} \lambda_{Q, -Q, O_{q} \pa^J O_{-q}} w^J , ~~~ w \gg 1 \ .
}
From \asymptotic\ we see that the LHS can be reproduced if the three-point couplings $ \lambda_{Q, -Q, O_{q} \pa^J O_{-q}} $ take the following value at large spin
\eqn\roughlyEq{
\lim_{J \gg 1} \lim_{Q \gg 1} \lambda_{Q, -Q, O_{q} \pa^J O_{-q}} = c_02^{\Delta_q -1} Q^{\Delta_q} {(\alpha \sqrt Q  q)^{J} \over J!} {J^{{5 \over 4} - \Delta_q} \Gamma(\Delta_q) \pi^{{1 \over 4}}} \ .
}
This is in accord with expectations from the EFT. Indeed, the operators $O_{q} \pa^J O_{-q}$ could be represented in EFT schematically as $|\pa \chi|^{\Delta_q} e^{i q \chi} \ \pa^J \ |\pa \chi|^{\Delta_q} e^{- i q \chi}$, see \lightopGB. The leading contribution at large $Q$ comes when all $J$ derivatives act on the exponential factor. This brings a factor of $q \mu \simeq \alpha \sqrt Q q$ in agreement with \roughlyEq. Let us emphasize again that \roughlyEq\ is different from the usual light-cone bootstrap, where spin $J$ and not the charge $Q$ is the largest parameter in the problem. It would be interesting to understand the interpolation between these two regimes.

%%%%%%%%%%%%%%%%%%%%%%%%%%%%%%%%%%%%%%%%%%%%%%%%%%%%%%%%%%%%%%%%%%%%

\newsec{Conclusions and Future Directions}

In this paper, we studied the four-point function of charged operators \fourpoint. Physically, it describes a correlation function of two light charged operators ${\cal O}_{-q} {\cal O}_{q}$ in a nontrivial state created by a heavy operator with large charge $Q \gg 1$ and scaling dimension $\Delta(Q) \gg 1$.

Our analysis is motivated by the results of \HellermanNRA\ and \MoninJMO\ which allow one to systematically compute \fourpoint\ in some $d=3$ CFTs with $U(1)$ symmetry using the Goldstone EFT. The result for the first few nontrivial orders is given by \EFTfourGB. We explained in detail how crossing is satisfied in this example and set up the corresponding bootstrap problem. One feature of the analysis is that the short distance OPE, $t$-channel in our notations, between the light operators  ${\cal O}_{-q} {\cal O}_{q}$ is not a part of the EFT. Another feature is that the $s$- and $u$- OPE channels do not have overlapping regions of convergence within the EFT approximation.

We noted that the short distance limit of the correlator in the large $Q$ limit is controlled by the infinite volume, macroscopic limit of correlators \macroonecross. This is a completely generic limit of correlation functions in CFTs, discussed recently in \LashkariVGJ, which we discussed in section 5.

Combining the macroscopic limit with the $s$- and $u$-channel OPEs in a way similar to \PappadopuloJK\ we obtained the system of equations for the three-point functions and anomalous dimensions \analyticequations. In deriving these equations, we assumed that only a finite number of Regge trajectories contribute to the correlator at the leading nontrivial order. This is a more abstract CFT analog of the statement that a putative EFT has a finite number of fields.

We proceeded by finding all solutions to this system of equations. The dimensions of operators as functions of spin are encoded in the roots of a certain polynomial \epssolpol. The three-point functions are then given by \dsol.

For one Regge trajectory the solution is unique and coincides with the results of the Goldstone EFT. By considering external states with spin (section 7), one can also show in this case that the spectrum is additive. For more than one trajectory there are more solutions. Some of them could be interpreted as an addition of extra particles to the EFT, while others do not look like something coming from a weakly coupled Lagrangian. Finding all solutions to the bootstrap problem formulated above comprises the main result of our paper.

There are many open questions that our analysis did not address. It seems that the most urgent one is to understand which solutions among the ones we found can be promoted to full-fledged solutions of crossing at all orders in ${1 \over Q}$ and thus could appear in actual CFTs. This would require analysis of the crossing equation at next orders in ${1 \over Q}$. Ideally, in this way we might hope to classify all possible ``phases'' that describe large $Q$ limits of CFTs. We also focused our attention on the case of $U(1)$ symmetry. This could be easily generalized.

An important part of our analysis is the macroscopic limit of the correlators \macroonecross. This is interesting on its own and deserves further investigation. It is clear that the thermodynamic limit of \LashkariVGJ\ does not always exist. This appears to be related to the existence of a nontrivial moduli space of vacua, as we briefly explained using the example of a free complex scalar. It would be extremely interesting to generalize the analysis of the present case to the more nontrivial scenarios of \HellermanVEG, \HellermanSUR. Another direction would be to consider generic heavy states, as in \LashkariVGJ,  and try to understand how universal hydrodynamic properties emerge, see, e.g., \HartnollAPF.

We also assumed that the spectrum of lightest operators of large charge is sparse. This assumption is connected to recent discussions of weak gravity conjecture \ArkaniHamedDZ. Indeed, if the lightest state of large charge $Q$ were an extremal Reissner-Nordstr{\"o}m black hole in an AdS gravity dual description, then we would expect the density of the corresponding operators to be $e^{S_{BH}}$. Extremal RN black holes are believed to be unstable and it would be interesting to see if this could be seen directly from CFT. This would require a generalization of our analysis to the case of the large degeneracy of states. One simple thing that we can already point out is that if the gap above the lightest large charge operator goes to 0 at large $Q$, then the macroscopic limit implies that such a CFT in Minkowski space at finite chemical potential will have a nonzero entropy per unit volume at zero temperature. 

Even though the $t$-channel is not easily accessible in the EFT regime, we noted that the situation is better in the macroscopic limit. In particular, in section 7.3 we considered a specific limit $w \gg \bar w \gg 1$ and identified the usual double twist operators as those that are responsible for the leading asymptotic of the EFT correlator. It would be interesting to see if this could be promoted to a systematic ${1 \over J}$ expansion, as is the case in the usual light-cone bootstrap.

Another interesting result of \HellermanNRA, which we did not address, was the universal $Q^0$ correction to the scaling dimension $\Delta(Q)$ given by the one-loop contribution of the Goldstone field to the effective action. We have not reproduced this contribution in our analysis. The first time it contributes to the correlator is at order ${1 \over Q^2}$. Microscopically, it corresponds to the contribution of the level one descendant. From the bootstrap point of view, we expect this correction to be fixed by consistency of the solution at higher orders in ${1 \over Q}$.

Further, it would be great if one could develop a numerical bootstrap approach to the problem discussed here. The discussed peculiarities of the large $Q$ limit make it potentially hard. On the other hand, the results of \Hasenbusch,\BanerjeeFCX\ are very encouraging and show that perhaps $Q \sim O(1)$ is already large enough. It would be very interesting to explore it further.

More generally, existing bootstrap techniques, both numerical and analytical, are sensitive to the low twist part of the CFT spectrum \SimmonsDuffinWLQ. Therefore it would be interesting to develop new approaches that would allow us to probe large twist CFT operators. The present paper is a very simple example of this kind.

\newsecB{Acknowledgments}

We would like to thank T. Dumitrescu, A. Dymarsky, A. Gorsky, S. Hellerman, N. Lashkari, H. Liu, S. Rychkov, M. Yamazaki, M. Watanabe and X. Yin for useful discussions. The work of A.Z. is supported by NSF grant 1205550. The work of D.J. and B.M. was supported in part by NSFCAREER grant PHY-1352084.

%%%%%%%%%%%%%%%%%%%%%%%%%%%%%%%%%%%%%%%%%%%%%%%%%%%%%

\appendix{A}{Solving Smoothness Conditions}

In this section we will derive all solutions of \analyticeq. It is convenient to separate the equations in \analyticeq\ into groups of $N$ equations, namely the equations $1$ to $N$, $2$ to $N+1$, $3$ to $N+2$ and so on. Each of these systems determines $d_i(z)$. To write this more concisely, let us define $N \times N$ matrices $V_n$ through recursion relations
\eqn\vanderrecurs{
V_{n+1} = V_n {\cal E}, \qquad n = 0,1,2, \dots \ ,
}
where $V_0$ is Vandermonde matrix and ${\cal E}$ is diagonal
\eqn\vandermonde{
V_0(z) = \left(
        \matrix{
          1 & \dots & 1 \cr
          \e_1(z) & \dots & \e_N(z) \cr
           \dots & \dots & \dots \cr
          \e_1(z)^{N-1} & \dots & \e_N(z)^{N-1}
        }
      \right), \qquad
{\cal E}(z) = \left(
               \matrix{
                 \e_1(z) &  & 0 \cr
                  & \ddots &  \cr
                  0 &  & \e_N(z) \cr
               }
             \right) \ .
}
Also, define columns of $d_i(z)$ and $W_n(z)$
\eqn\lamwcol{
\Lambda(z) = \left(
                \matrix{
                    d_1(z) \cr
                    \dots      \cr
                    d_N(z)
                    }
            \right), \qquad
A_n(z) =  \left(
                \matrix{
                    W_n(z)  \cr
                    W_{n+1}(z)      \cr
                    \dots                       \cr
                    W_{n+N - 1}(z)
                    }
            \right) \ .
}
Then the equations \analyticeq\ become
\eqn\smoothmat{
V_0 \Lambda = A_0, \quad V_1 \Lambda = A_1, \quad  V_2 \Lambda = A_2, \quad  \dots \ .
}
Each of these matrix equations is a linear system of $N$ equations for $N$ unknowns $\Lambda$. The determinants of the matrices $V_n$ are not zero in the physical regime of real $z>\sqrt{3}$
\eqn\detV{
\det V_n = \left( \prod_{i=1}^{N} \e_i \right)^{n} \prod_{i<j}\left( \e_j - \e_i \right) \neq 0
}
since we assumed that $\e_i \neq 0$ and $\e_i \neq \e_j$ for $i \neq j$. Thus, we can invert the matrices $V_n$ and solve {\smoothmat} to find $\Lambda$
\eqn\Lamsol{
\Lambda = V_0^{-1} A_0 = V_1^{-1} A_1 = V_2^{-1} A_2 = \dots \ .
}
Using the relations {\vanderrecurs}, we can write this as
\eqn\solone{\eqalign{
&\Lambda = V_0^{-1} A_0 \ , \cr
& (V_0 {\cal E} V_0^{-1}) A_0 = A_1 \ ,\cr
& (V_0 {\cal E} V_0^{-1}) A_1 = A_2 \ ,\cr
&(V_0 {\cal E} V_0^{-1}) A_2 = A_3 \  , \cr
&\dots
} }
Here, the first equation gives the solution for $d_i$ in terms of $\e_i, W_n$ and the rest of the equations give an infinite number of constraints on $\e_i, W_n$.

The matrix $V_0 {\cal E} V_0^{-1}$ can be directly computed and is given by
\eqn\VEVmatrix{
V_0 {\cal E} V_0^{-1} = \left( \matrix{
                          0    & 1 & 0 &  \dots & 0     \cr
                          0    & 0 & 1 &  \dots       & 0     \cr
                          \vdots & \vdots  & \vdots  & & \vdots         \cr
                          0 &0 &0 & & 1                 \cr
                          (-1)^{N-1} e_N(z) & (-1)^{N-2} e_{N-1}(z) & (-1)^{N-3}e_{N-2} & \dots   & e_1(z)
                                    }
                 \right) \ ,
}
where $e_k(z)$ are symmetric polynomials in $\e_i(z)$
\eqn\ekdef{
e_1(z) = \sum_i \e_i(z), \quad e_2(z) = \sum_{i<j} \e_i(z) \e_j(z) ,\quad  \dots, \quad e_N(z) = \prod_i \e_i(z) \ .
}
Only the last row of \VEVmatrix\ gives nontrivial equations in \solone. These are
\eqn\entwo{
\left(
    \matrix{
        W_0 & \dots & W_{N-1} \cr
       & \vdots &  \cr
        W_{N-1}  & \dots & W_{2(N-1)} \cr
          &\vdots   &
                                }
\right)
\left(
    \matrix{
   (-1)^{N-1} e_N \cr
    \vdots \cr
    e_1
    }
\right) =
\left(
    \matrix{
     W_N \cr
     \vdots \cr
     W_{2N-1}  \cr
     \vdots
    }
\right) \ ,
}
where we omitted the dependence on $z$ for the sake of brevity. A simple solution of the overdetermined linear system \entwo\ is for $e_k(z)$ to be a polynomial of order $k$
\eqn\esol{
e_k(z) = {\cal P}_k(z) = \sum_{n=0}^k a_n^{(k)} z^n  ,\quad k = 1,\dots, N \ .
}
Indeed, taking independent parameters to be the coefficients of polynomials $W_0, \dots, W_{N-1}$, ${\cal P}_1, \dots, {\cal P}_N$, the first equation in \entwo\ defines $W_N$, the second equation in \entwo\ defines $W_{N+1}$, etc.

Solving the first $N$ equations in \entwo\ we find that $e_k(z)$ is a rational function of $z$. As we prove below, to satisfy the rest of the equations it cannot have poles, so that \esol\ is the only solution of \entwo.

Since $e_k(z)$ are symmetric polynomials \ekdef, by Vieta's formula the scaling dimensions $\e_i(z)$ are given by $N$ different solutions of the $N$th order algebraic equation
\eqn\epssolpol{
\prod_{i=1}^N (x - \e_i(z)) = x^N - {\cal P}_1(z) x^{N-1} + {\cal P}_2(z) x^{N-2} + \dots + (-1)^N {\cal P}_N(z) =0 \ .
}
Finally, the three-point functions $d_i(z)$ are given by the first equation in \solone. Using Cramer's rule, we find
\eqn\dsol{
d_k(z) =  \prod_{i<j} (\eps_j(z) - \eps_i(z))^{-1} \det \left(  \matrix{
      1 & \dots & W_0 & \dots & 1 \cr
      \e_1(z) & \dots & W_1(z) & \dots & \e_N(z) \cr
        & \dots  & \dots & \dots  &  \cr
        \e_1(z)^{N-1} & \dots & W_{N-1}(z) & \dots & \e_N(z)^{N-1}
        } \right) \ ,
}
where the $k$-th column is made out of polynomials $W_n(z)$. The equations \epssolpol, \dsol\ give a complete solution of the smoothness conditions \analyticeq\ for spins $J \geq 2$.

Now let us show that \epssolpol, \dsol\ is the only solution. We need to show that $e_k(z)$ cannot have poles. Equivalently, we need to show that
\eqn\tracee{
\Tr\ {\cal E}^n = \sum_{i=1}^N \e_i(z)^n
}
does not have poles in $z$ for any positive integer $n$. Consider the equations \solone. They can be combined into matrix equations
\eqn\linsystother{\eqalign{
&(V_0 {\cal E} V_0^{-1})M_n = M_{n+1}, \qquad n = 0,1,2,\dots \ , \cr
&M_n = (A_n, \dots, A_{n+N-1}) \ ,
}}
and are solved by
\eqn\Esol{
V_0 {\cal E} V_0^{-1} = M_1 M_0^{-1} = M_2 M_1^{-1} = \dots \ .
}
Taking the trace of the $n$-th power of these equations we have
\eqn\trn{
\Tr\ {\cal E}^n = \Tr ( M_1 M_0^{-1})^n \ .
}
On the other hand, the equations \Esol\ require that $M_{n}M_0^{-1} = ( M_1 M_0^{-1})^n  $ and
\eqn\trnn{
\Tr\ {\cal E}^n = \Tr M_{n}M_0^{-1} \ .
}
Suppose that one of the $\eps_i(z)$ develops a singularity at $z_0$. This singularity must be consistent with \trnn . Let us expand \trnn\ around $z_0$. The crucial point is that a potential singularity on the RHS of \trnn\ could only come from $\det M_0 = 0$ and its maximal order does not depend on $n$.\foot{The actual behavior depends on the behavior of $\Tr M_{n}M_0^{-1} $, but it cannot be more singular than $1 \over \det M_0$.} The LHS of \trnn, which is equal to \tracee, has a singularity $\eps_i(z_0)^n$, whose strength is unbounded in contrast with the RHS of \trnn\ (it becomes more and more singular as $n$ grows). To reconcile these two facts, other $\eps_j(z)$ should soften $\eps_i(z_0)^n$ for large enough $n$. Imposing this cancelation it is trivial to see that for a finite number of Regge trajectories $N$ it is not possible for every $n$. Without loss of generality assume that close to $z_0$ we have $\eps_i(z) = c_i (z-z_0)^{- \alpha} + ...$. The conditions for cancelation of the singularity become
\eqn\condt{
\sum_{i=1}^{N} c_{i}^n = 0 , ~~~ n = n_0 , n_0 + 1 , ... \  ,
}
where $c_i$ are complex number (analogs of residues), which control the behavior of $\eps_i(z)$ near the singularity. The only solution to \condt\ is
\eqn\solC{
c_i = 0 \ .
}
Indeed, consider \condt\ as a linear system of a Vandermonde matrix of $c_i$'s acting on $(1,\dots,1)^T$. If $c_i \neq c_j$ and $c_i \neq 0$ the determinant of Vandermonde is non-zero and the system is inconsistent. If $c_i = c_j$ for some $i,j$, the system \condt\ can be reduced to a similar one with smaller $N$. Thus, we conclude that the only solution is \solC. Therefore, the assumed singularity was absent in the first place.

In the derivation above we tacitly assumed that $\det M_n(z)$ is not identically zero for all $n$ and $z$. Suppose one of them is identically zero. Without loss of generality assume $\det M_0 \equiv 0$. This implies
\eqn\detzeroM{
\det M_0 = \det (A_0, \dots, A_{N-1}) = \det (A_0, V_0 {\cal E} V_0^{-1} A_0 , \dots, (V_0 {\cal E} V_0^{-1})^{N-1} A_0 ) \equiv 0 \ .
}
Therefore, there must exist non-zero $\lambda_k(z)$ such that
\eqn\lambdalin{
\sum_{k=0}^{N-1} \lambda_k (V_0 {\cal E} V_0^{-1})^{k} A_0 \equiv 0 \ .
}
Using the first equation in \solone\ and \lamwcol , we can write \lambdalin\ as
\eqn\rewrite{
\sum_{k=0}^{N-1} \lambda_k(z) \e_i(z)^k d_i(z) \equiv 0, \qquad i = 1,\dots,N \ .
}
Consider this equation in the physical regime $z \geq \sqrt{3}$. We can divide by $d_i(z)>0$ since these are nonzero in the physical region. Further, since $\e_i(z) \neq \e_j(z)$ for $i\neq j$, the determinant of the system \rewrite\ is non-zero. Indeed, it is given by the Vandermonde determinant made of $\e_i(z)$'s. Consequently, the system \rewrite\ has only trivial solution $\lambda_k(z) \equiv 0$. Thus, physical constraints rule out the possibility that $\det M_n(z) \equiv 0$.

\listrefs

\bye